%% file: icse.tex
\documentclass[sigconf, screen=true]{acmart}

\input{Sections/packages.tex}

\newcommand{\interviewee}[1]{#1\xspace}
\newcommand{\sorin}{\interviewee{I1}}

\newcommand{\mathias}{\interviewee{I3}}

\newcommand{\davide}{\interviewee{I5}}

\newcommand{\navarro}{\interviewee{I7}}
\newcommand{\norbert}{\interviewee{I8}}

\newcommand{\per}{\interviewee{I14}}

\newcommand{\thilo}{\interviewee{I16}}
\newcommand{\enrique}{\interviewee{I17}}

\newcommand{\intindent}{.0cm}

\newcommand{\qmathias}[1]{\vspace{.00cm}\hangindent=\intindent \mathias: \textit{``#1''}}  
\newcommand{\qsorin}[1]{\vspace{.00cm}\hangindent=\intindent \sorin: \textit{``#1''}}  
\newcommand{\qdavide}[1]{\vspace{.00cm}\hangindent=\intindent \davide: \textit{``#1''}}  
\newcommand{\qnavarro}[1]{\vspace{.00cm}\hangindent=\intindent \navarro: \textit{``#1''}}  
\newcommand{\qnorbert}[1]{\vspace{.00cm}\hangindent=\intindent \norbert: \textit{``#1''}}  
\newcommand{\qper}[1]{\vspace{.00cm}\hangindent=\intindent \per: \textit{``#1''}}  
\newcommand{\qthilo}[1]{\vspace{.00cm}\hangindent=\intindent \thilo: \textit{``#1''}}  
\newcommand{\qenrique}[1]{\vspace{.00cm}\hangindent=\intindent \enrique: \textit{``#1''}}  

\newcommand{\scampusano}{\interviewee{S1}}  
\newcommand{\scamorga}{\interviewee{S2}} 
\newcommand{\sbrad}{\interviewee{S3}} 
\newcommand{\sdavid}{\interviewee{S5}} 
\newcommand{\sjulian}{\interviewee{S6}} 
\newcommand{\sanonmaturity}{\interviewee{S7}} 
\newcommand{\sanonmaturitytwo}{\interviewee{S8}} 
\newcommand{\sgeorg}{\interviewee{S10}} 
\newcommand{\stahsincankose}{\interviewee{S11}} 
\newcommand{\singo}{\interviewee{S12}} 
\newcommand{\sivan}{\interviewee{S13}} 
\newcommand{\sandrew}{\interviewee{S14}}  

\newcommand{\qscampusano}[1]{\vspace{.00cm}\hangindent=\intindent \scampusano: \textit{``#1''}}
\newcommand{\qscamorga}[1]{\vspace{.00cm}\hangindent=\intindent \scamorga: \textit{``#1''}}
\newcommand{\qsbrad}[1]{\vspace{.00cm}\hangindent=\intindent \sbrad: \textit{``#1''}}

\newcommand{\qsdavid}[1]{\vspace{.00cm}\hangindent=\intindent \sdavid: \textit{``#1''}}
\newcommand{\qsjulian}[1]{\vspace{.00cm}\hangindent=\intindent \sjulian: \textit{``#1''}}
\newcommand{\qsanonmaturity}[1]{\vspace{.00cm}\hangindent=\intindent \sanonmaturity: \textit{``#1''}}
\newcommand{\qsanonmaturitytwo}[1]{\vspace{.00cm}\hangindent=\intindent \sanonmaturitytwo: \textit{``#1''}}

\newcommand{\qsgeorg}[1]{\vspace{.00cm}\hangindent=\intindent \sgeorg: \textit{``#1''}}
\newcommand{\qstahsincankose}[1]{\vspace{.00cm}\hangindent=\intindent \stahsincankose: \textit{``#1''}}
\newcommand{\qsingo}[1]{\vspace{.00cm}\hangindent=\intindent \singo: \textit{``#1''}}
\newcommand{\qsivan}[1]{\vspace{.00cm}\hangindent=\intindent \sivan: \textit{``#1''}}
\newcommand{\qsandrew}[1]{\vspace{.00cm}\hangindent=\intindent \sandrew: \textit{``#1''}}

\setcopyright{acmcopyright}
\acmPrice{15.00}
\acmDOI{10.1145/3368089.3409743}
\acmYear{2020}
\copyrightyear{2020}
\acmSubmissionID{fse20main-p542-p}
\acmISBN{978-1-4503-7043-1/20/11}
\acmConference[ESEC/FSE '20]{Proceedings of the 28th ACM Joint European Software Engineering Conference and Symposium on the Foundations of Software Engineering}{November 8--13, 2020}{Virtual Event, USA}
\acmBooktitle{Proceedings of the 28th ACM Joint European Software Engineering Conference and Symposium on the Foundations of Software Engineering (ESEC/FSE '20), November 8--13, 2020, Virtual Event, USA}

\begin{document}
\title{Robotics Software Engineering: A Perspective from the Service Robotics Domain}

\author{Sergio Garc\'{i}a}
\affiliation{%
	\institution{University of Gothenburg $|$ Chalmers,\\ Gothenburg, Sweden}
}
\email{sergio.garcia@gu.se}

\author{Daniel Str{\"u}ber}
\affiliation{%
	\institution{Radboud University Nijmegen,\\ Nijmegen, Netherlands}
}
\email{d.strueber@cs.ru.nl}

\author{Davide Brugali}
\affiliation{%
	\institution{University of Bergamo,\\ Bergamo, Italy}
}
\email{davide.brugali@unibg.it}

\author{Thorsten Berger}
\affiliation{%
	\institution{University of Gothenburg $|$ Chalmers,\\ Gothenburg, Sweden}
}
\email{thorsten.berger@gu.se}

\author{Patrizio Pelliccione}
\affiliation{%
	\institution{University of Gothenburg $|$ Chalmers,\\ Gothenburg, Sweden  \\ and University of L'Aquila, \\ L'Aquila, Italy}
}
\email{patrizio.pelliccione@gu.se}


\begin{abstract}
	\looseness=-1
Robots that support humans by performing useful tasks (a.k.a., service robots) are booming worldwide.
In contrast to industrial robots, the development of service robots comes with severe software engineering challenges, since they require high levels of robustness and autonomy to operate in highly heterogeneous environments.
As a domain with critical safety implications, service robotics faces a need for sound software development practices.
In this paper, we present the first large-scale empirical study to assess the state of the art and practice of robotics software engineering. 
We conducted 18 semi-structured interviews with industrial practitioners working in 15 companies from 9 different countries and a survey with 156  respondents from 26 countries from the robotics domain. 
Our results provide a comprehensive picture of (i) the practices applied by robotics industrial and academic practitioners, including processes, paradigms, languages, tools, frameworks, and reuse practices, (ii) the distinguishing characteristics of robotics software engineering, and (iii) recurrent challenges usually faced, together with adopted solutions.
The paper concludes by discussing 
observations, derived hypotheses, and proposed actions for researchers and practitioners.
\end{abstract}

\ccsdesc[500]{Computer systems organization~Robotics}
\ccsdesc[500]{Software and its engineering}

\keywords{robotics software engineering,
	interviews,
	online survey}

\maketitle

\vspace{-0.25cm}
\input{Sections/intro.tex}

\input{Sections/background.tex}

\input{Sections/method.tex}

\input{Sections/rq1.tex}

\input{Sections/rq2.tex}

\input{Sections/rq3.tex}

\input{Sections/recommendations.tex}

\input{Sections/validity}

\input{Sections/conclusion-rec.tex}

\vspace{-0.15cm}
\looseness=-1
\section*{Acknowledgments}
Research supported by the EU H2020 Prog. under GA No. 731869 (Co4Robots) and the Centre of EXcellence on Connected, Geo-Localized, and Cybersecure Vehicle (EX-Emerge), funded by Italian Government under CIPE resolution n. 70/2017 (Aug. 7, 2017).
We are thankful to the participants for their collaboration.

\balance

\newpage

\bibliographystyle{ACM-Reference-Format}
\bibliography{Sections/sigproc}

\end{document}

%% file: Sections/packages.tex
\usepackage{pbox}

\usepackage{dirtytalk}

\usepackage{framed,lipsum}
\usepackage{microtype}
\usepackage[font=footnotesize]{caption}
\usepackage{booktabs} 
\usepackage{centernot}
\usepackage[Algorithmus]{algorithm}
\usepackage{algorithmic}
\usepackage{amsmath,amssymb,amsfonts}
\usepackage{balance}
\usepackage{fixltx2e}

\usepackage{graphicx}
\usepackage{subcaption}

\usepackage{array}
\usepackage{color}
\usepackage{listings}
\usepackage{amsmath,amssymb,amsfonts}
\usepackage{algorithmic}
\usepackage{graphicx}
\usepackage{textcomp}
\usepackage{xcolor}

\lstdefinelanguage{Xtext}{
	morekeywords={grammar, with, hidden, generate, as, import, returns, current, terminal, enum},
	keywordstyle=[2]{\textbf},
	morecomment=[l]{//}, 
	morecomment=[s]{/*}{*/}, 
	morestring=[b]",
	tabsize=4}

\usepackage{booktabs}
\usepackage{url}
\usepackage{microtype}

\newcommand{\quotebox}[3]{\vspace{.5em}\noindent\begin{tikzpicture}
	\node[align=center,draw,thin,minimum width=\columnwidth,inner sep=2.2mm] (titlebox)%
	{\parbox{0.95\columnwidth}{\hspace{-0.1cm}\noindent\textit{#2}}};\node[fill=white] (W) at ([xshift=#3]titlebox.south) {\small #1};%
	\end{tikzpicture}}

\newcounter{observationno}
\newcommand{\observation}[1]{\refstepcounter{observationno}Observation \the\numexpr\value{observationno}~(#1)}

\definecolor{codegreen}{rgb}{0,0.6,0}
\definecolor{codegray}{rgb}{0.5,0.5,0.5}
\definecolor{codepurple}{rgb}{0.58,0,0.82}
\definecolor{backcolour}{rgb}{0.95,0.95,0.92}
\definecolor{xtext}{RGB}{178, 14, 140}
\lstdefinestyle{mystyle}{
	commentstyle=\color{codegreen},
	keywordstyle=\color{magenta},
	numberstyle=\tiny\color{codegray},
	stringstyle=\color{codepurple},
	basicstyle=\scriptsize,
	breakatwhitespace=false,         
	breaklines=true,                 
	captionpos=b,                    
	keepspaces=true,                 
	numbers=none,                    
	numbersep=5pt,                  
	showspaces=false,                
	showstringspaces=false,
	showtabs=false,                  
	tabsize=2
}
\usepackage{threeparttable}

\usepackage{url}

\newcommand{\foot}[1]{\footnote{\url{#1}}}

\usepackage{xcolor}

\usepackage{centernot}
\usepackage{algorithm}
\usepackage{algorithmic}
\usepackage{amsmath,amssymb,amsfonts}

\usepackage[inline]{enumitem}
\usepackage{rotating}
\usepackage{multirow}
\usepackage{tikz}
\usepackage{tabularx}
\usepackage{tabulary}
\usepackage{threeparttable}
\usepackage{rotating} 
\usepackage{microtype}
\usepackage{url}
\usepackage[normalem]{ulem}
\usepackage{chngpage}
\usepackage{relsize}
\usepackage{xspace}
\usepackage[skins,breakable]{tcolorbox}

\usepackage{footmisc}

\newtcolorbox{mybox}[3][]
{
  colframe = #2!25,
  colback  = #2!10,
  coltitle = #2!20!black,  
  title    = {#3},
  #1,
}

\usepackage[colorinlistoftodos,prependcaption,textsize=tiny]{todonotes}

\newboolean{showcomments}
\setboolean{showcomments}{true} 
\ifthenelse{\boolean{showcomments}}
{\newcommand{\nb}[2]{
		\fcolorbox{black}{yellow}{\bfseries\sffamily\scriptsize#1}
		{\sf\small$\blacktriangleright$\textit{#2}$\blacktriangleleft$}
	}
	
}
{\newcommand{\nb}[2]{}
	
}

\usepackage[normalem]{ulem} 
\usepackage{xcolor}

\newcommand\parhead[1]{\vspace{.26mm}\noindent\textbf{{#1}}.}

\newcommand{\secref}[1]{Sec.\,\ref{#1}}

\newcommand{\Figref}[1]{Figure\,\ref{#1}} 
\newcommand{\tabref}[1]{Table\,\ref{#1}}

\usepackage{graphicx}

\lstdefinelanguage{LTL}{
	otherkeywords={l1, l2, l3, l4, l5, chargingdock, raise_alarm, request_help, charge_battery, grasp_object, office1, office2},
	morekeywords={>,<,)},
	morecomment=[l]{.}
	sensitive=true}

\lstdefinelanguage{Xtext}{
	morekeywords={grammar, with, hidden, generate, as, import, returns, current, terminal, enum},
	keywordstyle=[2]{\textbf},
	morekeywords=[3]{name, String, Integer, LTL_formula},
	keywordstyle=[3]\color{xtextOrange},
	morecomment=[l]{//}, 
	morecomment=[s]{/*}{*/}, 
	morestring=[b]",
	morestring=[b]',
	moredelim=**[is][\color{gray}]{`}{`},
	tabsize=4}

\definecolor{xtextOrange}{RGB}{171,48,0}
\definecolor{xtextBlue}{RGB}{42,8,255}

\makeatletter
\newcommand\footnoteref[1]{\protected@xdef\@thefnmark{\ref{#1}}\@footnotemark}
\makeatother

\usepackage{array}
\newcolumntype{P}[1]{>{\centering\arraybackslash}p{#1}}

%% file: Sections/intro.tex
\section{Introduction}
\label{sec:intro}



\looseness=-1
Service robots are a rising robotics domain with broad applications in many fields, such as logistics, healthcare, telepresence, maintenance, domestic tasks, education, and entertainment.
A service robot is ``{\em a type of robot that performs useful tasks for humans or equipment excluding industrial automation applications}''\,\cite{iso}.
Service robots are increasingly becoming part of our lives~\cite{mar}, being the service robotics market estimated to reach a value of \$ 102.5 billion by 2025\,\cite{markets17}.
Compared to industrial robotics, the service robotics domain is more challenging, since these robots usually operate in unconstrained environments, often populated by humans, requiring high degrees of robustness and autonomy.

\looseness=-1
Despite software playing an ever-increasing role in robotics, the current software engineering (SE) practices are perceived as insufficient, often leading to error-prone and hardly maintainable and evolvable software. 
Robotic systems are an advanced type of cyber-physical system (CPS), made up of an intricate blend of hardware, software, and environmental components.
SE, despite its beneficial role in other CPS domains (e.g., automotive, aeronautics), has traditionally been considered an auxiliary concern of robotic system construction\,\cite{brugali2009software}. 
A possible reason is that robots in factory automation have built-in proprietary controllers for repetitive tasks, therefore, allowing a simple programming style. 
The heavy lifting is in the domains of mechanics, electronics, and automatic control\,\cite{Brugali2005}. 
In contrast, to achieve autonomy when interacting in highly heterogeneous environments, service robots are equipped with a large variety of functionalities for perception, control, planning, learning, and multimodal interaction with the human operator.

The integration, customization, and evolution of these functionalities give rise to a large amount of complexity, the management of which is a challenging task.
SE systematic practices could play a crucial role in the management of such complexity, yet, according to a recent study\,\cite{bozhinoski2019safety}, the community of SE and robotics is still not consolidated.
In fact, deficiencies in the control systems of service robots stemming from a bad management of their inherent complexity are especially expensive if found after the system is put in operation.
For instance, in 2016, the Japan Aerospace Exploration Agency lost 286 million dollars due to an error in the control system of its flagship astronomical satellite.
The system commanded a thruster jet to fire in the wrong direction, resulting in ten critical parts breaking off the satellite body\,\cite{accident-satellite}. 


\looseness=-1
Our long-term goal is to establish effective SE practices for a substantial lifting of the state-of-practice for engineering service robotics and provide guidance for practitioners.
We also aim to help the research community to scope future research that meets industrial needs.
According to Bozhinoski et al., one of the challenges ``{\em is to promote a shift towards well-defined engineering approaches able to stimulate component supply-chains and significantly impact the robotics marketplace}"\,\cite{bozhinoski2019safety}.
However, systematic studies about the specific software development practices and tools applied in service robotics as well as the challenges faced by practitioners in this domain are currently lacking.
Academics and practitioners would benefit from such a study, as it would allow us to quantify and check existing subjective beliefs and anecdotal evidence~\cite{tichy2000hints}, and may contribute to the formation of a body of knowledge on SE in robotics. 
Practitioners would further benefit from such a study as companies need to make decisions about huge investments.
Having a snapshot of the current state-of-practice may help them in making informed decisions.
To these ends, this paper assesses the current SE practices applied to the domain of service robotics, as well as its distinguishing characteristics and faced challenges. 
Our research questions are:

\noindent{}\textbf{RQ1.} \textit{What practices are applied in SE for service robots?} 
We study used SE principles, including processes and quality assurance techniques; also, we investigate technical spaces in use, including frameworks, middleware, software languages, and tools.
Finally, we study reuse practices, including the frequency and motivations for reusing existing solutions 
or, instead, for favoring self-developed solutions.

\looseness=-1
\noindent{}\textbf{RQ2.} \textit{How is robotics SE perceived to be different from SE in other domains?}
We identify the special characteristics 
that make robotics SE stand out from SE in other domains.
We are interested in the effect of the robots' features, 
their hardware, their requirements on autonomy, and the heterogeneity of contexts where they are deployed.

\noindent{}\textbf{RQ3.} \textit{What challenges do practitioners face related to SE for service robots?}
We elicit the most common SE challenges faced by practitioners in service robotics, as well as the solutions applied.

\looseness=-1
To collect data, we conducted 18 semi-structured interviews with industrial robotics experts working for 15 companies from 9 different countries.
We accompany this study with an online survey, targeting industrial and academic practitioners in the robotics domain, from which we collect 156 responses.
Academic practitioners are those who work for an academic institution (we recruited most of our respondents from significant GitHub repositories with robotics-specific code).
Among the survey respondents, we identified practitioners working for 58 companies from 20 different countries and 16 academic institutions from 10 different countries.
To the best of our knowledge, our study is the first with this ambition.

\looseness=-1
In our study, we discovered that roboticists are predominantly focused on implementation and real-world testing, often favored over simulation. 
We learned that robotic control systems are typically developed as component-based systems, implemented by developers who may come from different backgrounds (e.g., mechanical, electrical, or software engineers).
We also elicited the main characteristics 
of robotics SE, where the cyber-physical nature of robots and the variety of disciplines required to develop a complete robotic system were highlighted.
These characteristics increase the complexity of robots' control software, calling for systematic practices as modeling and the usage of software architectures to improve the development process~\cite{bures2015software}. 
Our respondents ranked challenges related to robustness and validation as most pressing, and typically address them by applying thorough testing processes.
Based on our observations, we also identify research themes that deserve further investigation.
We provide conjectures for why these themes are currently under-investigated and recommendations for both researchers and practitioners.
%
In summary, we contribute:

\begin{itemize}[leftmargin=4mm]
	\item The findings from our interview data, including qualitative information about applied practices, distinguishing characteristics, and faced challenges and solutions in service robotics.
	\item The results from our online survey, which supports a quantitative assessment of the trends regarding SE practices in robotics.
	\item A replication package\,\cite{replication-package}
	containing 
	\begin{enumerate*}
		\item the questionnaire and the anonymized responses from our online survey 
		\item our interview guide and used open-ended questions,  
		\item the codebook from the qualitative data analysis,
		\item and our quantitative data analysis scripts.
	\end{enumerate*}
	\item Discussion of observations, derived hypotheses, and proposed actions for researchers and practitioners.
\end{itemize}

\vspace{-0.2cm}

%% file: Sections/background.tex
\section{Related Work}
\label{sec:background}

Current achievements in defining standardized solutions, reference architectures, and model-driven approaches for the development of reusable robotic systems are documented in various survey papers briefly summarized in the following.

\looseness=-1
Brugali et al.~\cite{Brugali2007} present five challenges that characterize  robotics, namely 
\begin{enumerate*}
	\item the identification of stable requirements of robotic systems,
	\item the definition of abstract models to cope with hardware and software heterogeneity,
	\item the seamless transition from prototype testing and debugging to real system implementation,
	\item the deployment of robotic applications in real-world environments, and 
	\item the design and integration of software control systems depending on highly heterogeneous technologies.
\end{enumerate*}
Brugali et al.~\cite{Brugali2007b, Brugali2010, brugali:2015:roboticsmde} have reviewed existing approaches that exploit component-based SE (CBSE) and Model-Driven Engineering (MDE) in robotics. Notable examples of component models and associated MDE toolchains specifically defined for developing robotic applications are OROCOS~\cite{Orocos2003}, SmartSoft~\cite{Schlegel2007}, BRICS~\cite{Bruyninckx2013}, and LAAS-BIP~\cite{Ingrand2011}.

\looseness=-1
MDE has also been applied recently in service robotics as a core technology to compose systems out of exchangeable components and move towards well-engineered system development processes as opposed to current craftsmanship practices~\cite{mar}. 
Steck et al.\,\cite{SLS11} propose a system that exploits design- and run-time models to build robotic systems and support them in their decision-making process.
Closely related to MDE are Domain-Specific Languages (DSLs), which have also been highlighted for their benefits, including the improvement of composability and system integration~\cite{mar}.
The research community has already worked in the last years on the use of DSLs for the development of robotic  systems\,\cite{Schmidt2006,SHLS09,SLS11,Bruyninckx2013}. 
Nordmann et al.\,\cite{Nordmann2016} present a survey of more than 130 papers on DSLs in robotics. 
The papers are classified according to robotic functional subdomains, such as Kinematics and Motion Control.

Elkady et al.~\cite{Elkady2012} present a literature survey and attribute-based bibliography of the state-of-the-art in robotic middleware frameworks. For example, the Middleware for Mobile Robot Applications (MIRO)~\cite{MIRO2002} provides a robotic hardware abstraction layer that simplifies the development of robot control applications. 

\looseness=-1
Ahmad et al.\,\cite{Ahmad2016} conducted a systematic mapping study on software architectures for robotic systems, which reveals that the conventional concepts of architectural frameworks and architectural notations are well integrated into robotics.
However, they also identified a  lack of focus on architecture-specific evaluation for validating requirements for robotic software.
In addition, Kortenkamp et al.\,\cite{kortenkamp2016robotic} present various approaches to
architecting robotic systems with an emphasis on programming tools and
environments.  

Ingrand et al.\,\cite{ingrand2017deliberation} provide a survey that introduces a global perspective of autonomous robots' deliberation---i.e., endowing robotic systems with adaptable and robust functionalities.

\looseness=-1
Estef\'{o} et al.~\cite{estefo2019robot} interviewed and surveyed Robot Operating System (ROS)~\cite{quigley2009ros} developers.
They report on the issues related to resources reuse within this ecosystem. 

\looseness=-1
Bozhinoski et al.~\cite{bozhinoski2019safety} provide a systematic mapping study on the state of the art in safety for mobile robotic systems. 
Focusing on an SE perspective, they classify safety solutions from 58 primary studies concerning trends,  characteristics, and industrial adoption.
They found that existing solutions are largely not ready to be used in uncontrollable and unknown, often shared with humans environments. 
The only source of information of this study consists of published papers that have been selected as primary studies. 
This motivates us to perform a study to better investigate these aspects and by involving industrial and academic practitioners in robotics.

\looseness=-1
Synergies with SE have been investigated also for other domains that share common characteristics with robotics.
Amershi et al.~\cite{Amershi2019} conducted a study observing software teams at Microsoft as they develop AI-based applications. 
Altinger's Doctoral Thesis~\cite{Altinger2017} reports on fault prediction techniques applied to automotive software development.
Bures et al.~\cite{Bures2017} report on challenges and promising solutions in the area of SE for Smart Cyber-Physical Systems.



%% file: Sections/method.tex
\section{Methodology}
\label{sec:method}

\begin{figure}[b]
	\vspace{-0.2cm}
	\centering
	\includegraphics[trim=0.5cm 3.7cm 0.5cm 4cm, clip=true, width=\columnwidth]{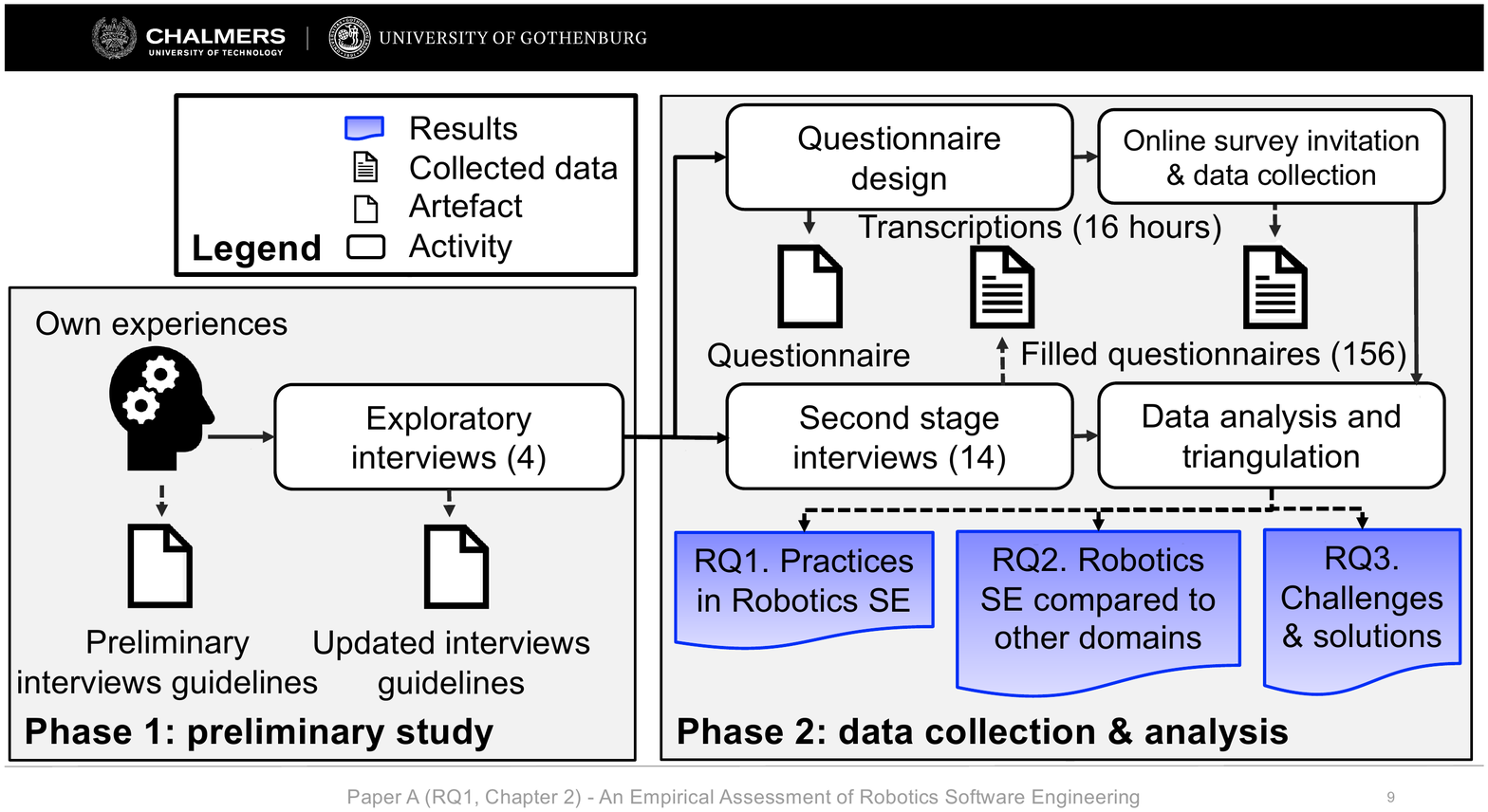}   
	\setlength{\abovecaptionskip}{-9pt} 
	\caption{Research methodology overview}
	\label{fig:method}
	\vspace{0.2cm}
\end{figure}

\looseness=-1
\Figref{fig:method} summarizes our methodology. For an initial set of four exploratory interviews, we relied on our own experiences as SE researchers in various robotics projects~\cite{co4robots, BRICS}
where we collaborated with robotics engineers, 
many of them from industry. 
In those interviews, we asked previous collaborators, who are expert robotics engineers, about their software engineering practices, tools, and challenges, following the typical software-engineering phases (from requirements via design over implementation and quality assurance), while also aiming to understand the organizational and business context. 
This helped us sharpening our knowledge of the typical terminology used in robotics.
Based on this knowledge and our research questions, we designed an interview guide and, in parallel, a survey questionnaire for the second phase of our study.
Our replication package\,\cite{replication-package} provides the questionnaire, an analysis script, detailed results, and our interview guide.

\subsection{Survey}\label{sec:survey}
\looseness=-1
\parhead{Design}
We implemented the survey questionnaire with Google Forms, designing it to take no more than 15--20 minutes to complete. It consists of 47 questions (27 optional, 20 mandatory) divided into the sections: robotic platforms and context (e.g., which types of robots are engineered), activities and paradigms (e.g., to which extent requirements engineering is performed or what tools are used), development practices (e.g., whether and how software artifacts are reused), quality assurance (e.g., where test data is obtained from), mission specification (e.g., whether the robot behavior is hard-coded), challenges and solutions (e.g., concerning applying AI techniques), general perception (e.g., what characteristics are unique to the domain), and finally a section on the respondents' demographics (e.g., years of experience). The majority of questions was closed-ended, using either checkboxes or 5-point Likert scales, the latter to express the frequencies of applying certain activities (e.g., code review), ranging from never via almost never, sometimes and very often, to always. We also allowed declaring ``Don't know.''


\looseness=-1
With this design, we followed an inductive and a deductive strategy: on the one hand (inductive), we used questions similar to the ones asked during the interviews so to generalize our findings; on the other hand (deductive), we asked about hypotheses (proposed explanations for phenomena) we learned about in the exploratory interviews. Some examples of such hypothesis are:
\begin{enumerate*}
	\item robotics software developers tend to ``reinvent the wheel'' instead of reusing already existing solutions; or
	\item some of the most important challenges in the field are the lack of standardized solutions and of internal and external documentation.
\end{enumerate*}
To this end, we designed questions such as: \textit{``Have you ever developed a software component from scratch rather than reusing an existing (either self-developed or third-party) one? If yes, why?''}
After the initial design, we tested the questionnaire internally among the authors, and five times externally with industrial and academic colleagues with expertise in robotics to fine-tune length, phrasing, and terminology before distribution.


\parhead{Distribution}
Our distribution strategy was three-fold. First, we used social media (Twitter and LinkedIn) to distribute our survey invitation, clearly asking for responses from practitioners in robotics. 
Social media stimulated the response rate, since it is inherently interactive and practitioners are active there---e.g., we knew examples from Twitter and Linkedin allowed searching for specific companies and roles. 
Second, we created a list of contacts from our previous projects and collaborations. Third, we mined GitHub repositories to extract their contributors' public email addresses. 
Specifically, we sought repositories from well-known tools (e.g., Gazebo~\cite{koenig2004design}), frameworks (e.g., ROS~\cite{quigley2009ros}, ROS2~\cite{maruyama2016exploring}, OROCOS~\cite{Orocos2003}, Yarp~\cite{metta2006yarp}), and libraries (e.g., OpenCV~\cite{pulli2012real}).
We also looked into repositories of well-known companies (e.g., Boston Dynamics), companies found in the media~\cite{construct},
and companies from survey respondents who provided their email addresses.
%
%
We also asked recipients to distribute our survey to other potential respondents.
Finally, to improve gender balance, we used Google to find female robotics engineers and found articles on this topic~\cite{women}.



\looseness=-1
\parhead{Survey Respondents}
We sent out 2499 invitations and collected 156 responses (response rate of 6\%, in line with the expectations of SE survey's response rate~\cite{shull2007guide}).
The most common occupation among the respondents is industrial practitioner with 60.9\% (31.4\% of leading technical roles and 29.5\% of programmers).
We also received a substantial number of responses from academic practitioners 
(57, 36.5\% of the total), so we decided to not filter them out and instead use that data to compare practices applied by academic and industrial practitioners.
The last group is formed by practitioners who did not identify themselves as industrial nor academic, and people who decided not to answer with their occupation (4, 2.6\%). 

We used provided email addresses to identify the companies for which our respondents work, and the associated countries.
That way, we mapped 72 of our industrial practitioners to 58 companies from 20 different countries, distributed over geographical areas as follows (number of respondents between parenthesis): 
\begin{enumerate*}[label=\alph*)]
	\item Europe (51), 
	\item North America (13), 
	\item Asia (7), and
	\item South America (1).
\end{enumerate*}
We also mapped 22 of our academic practitioners to 16 universities from 10 different countries from the following geographical regions: 
\begin{enumerate*}[label=\alph*)]
	\item Europe (18),
	\item Asia (1),
	\item South America (1), and
	\item Oceania (1).
\end{enumerate*}

\looseness=-1
Almost half of our respondents have between five and 10 years of experience in robotics (43.6\%), 17.3\% between 10 and 15, and 11.5\% more than 15.
Out of the total, 24.4\% of the respondents have between one and five years of experience and 3.2\% did not provide an answer to this question.
Regarding the size of the company, almost half of the answers (46.8\%) belong to respondents who work for companies with 50 or more employees.
Then, 11.7\%, 22.1\%, and 15.6\% of the respondents work for companies with 31-50, 11-30, and 1-10 employees, respectively.
Among all the respondents, 91.2\% are men, 6.1\% are women, and 2.7\% preferred not to specify.
In the remainder, we refer to individual survey respondents as S1, S2, and so on. 

\vspace{-12pt}

\subsection{Interviews}

\parhead{Design and Conduction} We designed an interview guide targeting interviews of around 45-60 minutes length. The guide started with questions regarding the interviewee's background (e.g., subject of university studies and previous roles at companies) and roughly resembled the questionnaire's structure and addressed each research question. The interviews were semi-structured: we had question blocks we wanted to cover, but allowed the interview to flow freely based on interesting aspects the interviewee brought up, where we then dug deeper if necessary. Before each interview, we also studied the survey responses to prepare specific questions, such as for clarification.
We also emphasized special topics depending on the interviewee and their role at the company. For instance, CEOs were usually able to provide much information regarding the company's organization or the overall project life-cycle. Software and robotics engineers mostly provided information on development issues, engineering paradigms, and technological spaces. Following a grounded-theory-like approach for the analysis (explained shortly), we also slightly changed our attention to frequent topics that arose along with the interviews, such as the dominant use of ROS as a robotics framework, its causes, and benefits and shortcomings.

\looseness=-1
\parhead{Interviewees}
We recruited our interviewees:
\begin{enumerate*}
	\item among the survey respondents who agreed to be contacted for follow-up questions,
	\item targeting domain experts using social media (see explanation in Sec.~\ref{sec:survey}), and
	\item among previous colleagues with expert robotics knowledge.
\end{enumerate*} 
We performed 18 interviews, summing up to 16 hours of interview data. 
As detailed in \tabref{table:participants}, the interviewees belong to 15 companies from nine different countries (16 interviewees work in Europe and two of them in North America), with
%
roles ranging from developer to CEO, giving us a broad perspective. 
We made sure that all interviewees are pure industrial practitioners.
The table also shows our interviewees' overall working experience within robotics and briefly characterizes each company.
In the remainder of the paper, we refer to individual interviewees with I1, I2, and so on.

%

\input{tables/participants2.tex}

\subsection{Data Analysis}
We analyzed the survey quantitatively by creating diagrams and manually aggregating responses to questions. To understand trends in the Likert-scale questions, we created bar charts (many of which are shown in the remainder). Furthermore, we analyzed the survey results qualitatively by inspecting responses to the open-ended questions. We created a technical report with the detailed results (available in our replication package\,\cite{replication-package}), also separating the results by the respondent's role, specifically: industrial practitioner (further sub-divided into leading technical role and programmer) and academic practitioner.
We make this separation to understand any deviations based on the role, and if there are, we report relevant differences separately. 
After creating the technical report, we triangulated the results with those from the interviews as follows.

We analyzed the interviews by transcribing them and then performing open coding\,\cite{strauss.ea:1990:opencoding}, as known from grounded theory\,\cite{corbin1990grounded}. Two authors coded the interviews iteratively, continuously discussing and refining the codes and organizing them in a hierarchy. We defined a codebook consisting of 373 codes, which we provide in our replication package\,\cite{replication-package} .
Based on the codes, we triangulated results from the survey and interviews (mapping codes to parts of the technical report about the survey data).
We report the results by first providing the context in which our survey respondents and interviewees perform their engineering, shortly in \secref{sec:context}.
Thereafter, in Sections \ref{sec:rq1}, \ref{sec:rq2}, and \ref{sec:rq3}, we use a narrative style to answer our research questions RQ1, RQ2, and RQ3, respectively.



\subsection{Robotic Platforms and Context}\label{sec:context}

\input{tables/service-field.tex}

\looseness=-1
Most respondents (75\%) affirm that they use (ground) mobile robots in their projects.
Collaborative robots and mobile manipulators come in second and third place with 45.5\% and 42.3\%, respectively.

\looseness=-1
Table~\ref{tab:service} gives an overview of the types of products and services provided by the respondents and what field they are applied to.
We find a predominance of planning modules (64.1\%) and low-level functionalities (63.5\%)  as the most common types.
Industrial practitioners tend to provide more complete robotic systems (59.6\%) than the academic ones (37.5\%).
Our survey allowed us to enter an open-ended text after answering ``Other,'' but despite the number of people who chose this option only 13 open-ended responses were collected for this question.
Examples of answers include testing systems, real-time operating systems, and high-level vision solutions.
The most common application fields are factory automation and research for service robots  (48.1\% for both).
The data shows a tendency among industrial practitioners towards factory automation (51.1\%) while academic practitioners favor research for service robotics (60.7\%).

%% file: tables/participants2.tex
\begin{table}[t]
	\centering
	\vspace{.9cm}
	\caption{List of interviewees}\label{table:participants}
	\vspace{-.3cm}
	\footnotesize
	\begin{threeparttable}
\begin{tabular}{ p{0.001cm}  p{0.005cm} p{0.25cm} p{0.25cm} p{6.2cm}  }
	\toprule
	&  \hspace{-0.2cm}\parbox[t]{0.2cm}{\textsf{\textsf{Cp.\textsuperscript{1}}}}  & \textsf{Exp.\textsuperscript{2}} & \textsf{Role\textsuperscript{3}} & \textsf{Company description}\\
	\midrule
	\hspace{-0.2cm}\parbox[t]{0.2cm}{\textsf{1}}	
	& 
	\hspace{-0.2cm}\parbox[t]{0.2cm}{\textsf{A}}
	& 6 & RSE & 
	Small, specialized in outdoor mobile robots and professional Research and Development (R\&D) consultancy.\\
	\hspace{-0.2cm}\parbox[t]{0.2cm}{\textsf{2}}	
	& \hspace{-0.2cm}\parbox[t]{0.2cm}{\textsf{B}} & 18 & CEO & 
	Small,  focuses on autonomous navigation solutions.\\
	\hspace{-0.2cm}\parbox[t]{0.2cm}{\textsf{3}}	
	& \hspace{-0.2cm}\parbox[t]{0.2cm}{\textsf{C}} & 5 &   PM & 
	Large, mainly works in the field of engineering and electronics. Currently investing in artificial intelligence and robotics.
	\\
	\hspace{-0.2cm}\parbox[t]{0.2cm}{\textsf{4}}	
	& \hspace{-0.2cm}\parbox[t]{0.2cm}{\textsf{C}} & 6  &  PM & 
	\\
	\midrule
	\hspace{-0.2cm}\parbox[t]{0.2cm}{\textsf{5}}	
	 & \hspace{-0.2cm}\parbox[t]{0.2cm}{\textsf{D}} & 15 &  RSE & 
	 Large technological center, mainly works in applied R\&D.
	 \\
	\hspace{-0.2cm}\parbox[t]{0.2cm}{\textsf{6}}	
	& \hspace{-0.2cm}\parbox[t]{0.2cm}{\textsf{E}} & 2  &  RSE & 
	Small, manufactures humanoid robots and robotic platforms. 
	\\
	\hspace{-0.2cm}\parbox[t]{0.2cm}{\textsf{7}}	
	& \hspace{-0.2cm}\parbox[t]{0.2cm}{\textsf{F}} & 8 &   RSE & 
	Small,   manufactures mobile service robotic platforms. 
	\\
	\hspace{-0.2cm}\parbox[t]{0.2cm}{\textsf{8}}	
	& \hspace{-0.2cm}\parbox[t]{0.2cm}{\textsf{G}} &5  &   RSE & 
	Large, works in the intralogistics sector.
	\\
	\hspace{-0.2cm}\parbox[t]{0.2cm}{\textsf{9}}	
	& \hspace{-0.2cm}\parbox[t]{0.2cm}{\textsf{H}} & 2.5  & RSE & 
	Technology-oriented large company, also invests in robotics. 
	\\
	\hspace{-0.2cm}\parbox[t]{0.2cm}{\textsf{10}}	
	& \hspace{-0.2cm}\parbox[t]{0.2cm}{\textsf{I}} & 6 &  RSE & 
	Large technology institute, which focuses in diverse industrial fields, including mobile robots.
	\\
	\hspace{-0.2cm}\parbox[t]{0.2cm}{\textsf{11}}	
	& \hspace{-0.2cm}\parbox[t]{0.2cm}{\textsf{C}} & 14 & RSE  &
	
	\\
	\hspace{-0.2cm}\parbox[t]{0.2cm}{\textsf{12}}	
	& \hspace{-0.2cm}\parbox[t]{0.2cm}{\textsf{J}} & 8  &  SAE& 
	Medium-sized, provides self-driving products and services.
	\\
	\hspace{-0.2cm}\parbox[t]{0.2cm}{\textsf{13}}	
	& \hspace{-0.2cm}\parbox[t]{0.2cm}{\textsf{K}} & 4  & CEO & 
	Micro-business, develops and rents customer-tailored robots.
	\\
	\hspace{-0.2cm}\parbox[t]{0.2cm}{\textsf{14}}	
	& \hspace{-0.2cm}\parbox[t]{0.2cm}{\textsf{L}} & 4 &  CEO & 
	Micro-business, develops and rents modular robot solutions.
	\\
	\hspace{-0.2cm}\parbox[t]{0.2cm}{\textsf{15}}	
	& \hspace{-0.2cm}\parbox[t]{0.2cm}{\textsf{M}} & 21 &  PM  & 
	Large,  mainly works in the aviation sector, but  also  with unmanned air vehicles.\\
	\hspace{-0.2cm}\parbox[t]{0.2cm}{\textsf{16}}	
	& \hspace{-0.2cm}\parbox[t]{0.2cm}{\textsf{N}} & 11 &   PM & 
	Large research institute, focuses on industrial projects like artificial intelligence in automation.
	\\
	\hspace{-0.2cm}\parbox[t]{0.2cm}{\textsf{17}}	
	& \hspace{-0.2cm}\parbox[t]{0.2cm}{\textsf{J}} & 7  &  RSE & 
	\\
	\hspace{-0.2cm}\parbox[t]{0.2cm}{\textsf{18}}	
	& \hspace{-0.2cm}\parbox[t]{0.2cm}{\textsf{O}} & 8 &  CTO  & 
	Small,  works in educational technology using service robots.\\
\bottomrule
\end{tabular}



		\textsuperscript{1}Company name (letter code), \textsuperscript{2}experience with robotics in years.
		\\
		\textsuperscript{3}{Roles:} 
		    Robotics Software Engineer (RSE);
			Chief Executive Officer (CEO);
			Project Manager (PM);
			Application Engineer (AE);
			Chief Technology Officer (CTO).	
	\end{threeparttable}
	\vspace{-0.5cm}
\end{table}

%% file: tables/service-field.tex
\begin{table}[b]
	\footnotesize
  	\vspace{0.4cm}
 	\caption{Provided service or product and application fields}
	\vspace{-0.3cm}
	\label{tab:service}
	\begin{tabular}{ p{2.6cm}  P{0.8cm} P{1.6cm} P{1.6cm} }
		\toprule
		& \textsf{Overall}  & \textsf{Industrial pract.} & \textsf{Academic pract.}\\
		\midrule
		
		\hspace{-0.2cm}\parbox[t]{2.6cm}{\textsf{Planning modules}}	
		&
		64.1\,\%
		&
		58.5\,\%
		&
		75\,\%
		\\
		
		\hspace{-0.2cm}\parbox[t]{2.6cm}{\textsf{Low-level functionalities}}
		&
		63.5\,\%
		&
		59.6\,\%
		& 
		71.4\,\%
		\\

		\hspace{-0.2cm}\parbox[t]{2.6cm}{\textsf{Complete system}}
		&
		51.3\,\%
		&
		59.6\,\%
		& 
		37.5\,\%
		\\

		\hspace{-0.2cm}\parbox[t]{2.6cm}{\textsf{Drivers}}
		&
		44.9\,\%
		&
		43.6\,\%
		& 
		44.6\,\%
		\\

		\hspace{-0.2cm}\parbox[t]{2.6cm}{\textsf{Other}}
		&
		9\,\%
		&
		9.6\,\%
		&
		8.9\,\%
		\\
		
		\midrule 
		
		\hspace{-0.2cm}\parbox[t]{2.6cm}{\textsf{Factory automation}}
		&
		48.1\,\%
		&
		51.1\,\%
		&
		42.9\,\%
		\\
		
		\hspace{-0.2cm}\parbox[t]{2.6cm}{\textsf{Service robotics research}}
		&
		48.1\,\%
		&
		38.3\,\%
		&
		60.7\,\%
		\\
		
		\hspace{-0.2cm}\parbox[t]{2.6cm}{\textsf{Others}}
		&
		34.6\,\%
		&
		40.4\,\%
		&
		26.8\,\%
		\\
		
		\hspace{-0.2cm}\parbox[t]{2.6cm}{\textsf{Transportation}}
		&
		24.4 \,\%
		&
		24.5\,\%
		&
		23.2\,\%
		\\
		\bottomrule
	\end{tabular}
\end{table}

%% file: Sections/rq1.tex
\begin{figure*}[ht!]
	\begin{minipage}[t]{0.32\textwidth}
		\includegraphics[
		clip=true, width=\linewidth]{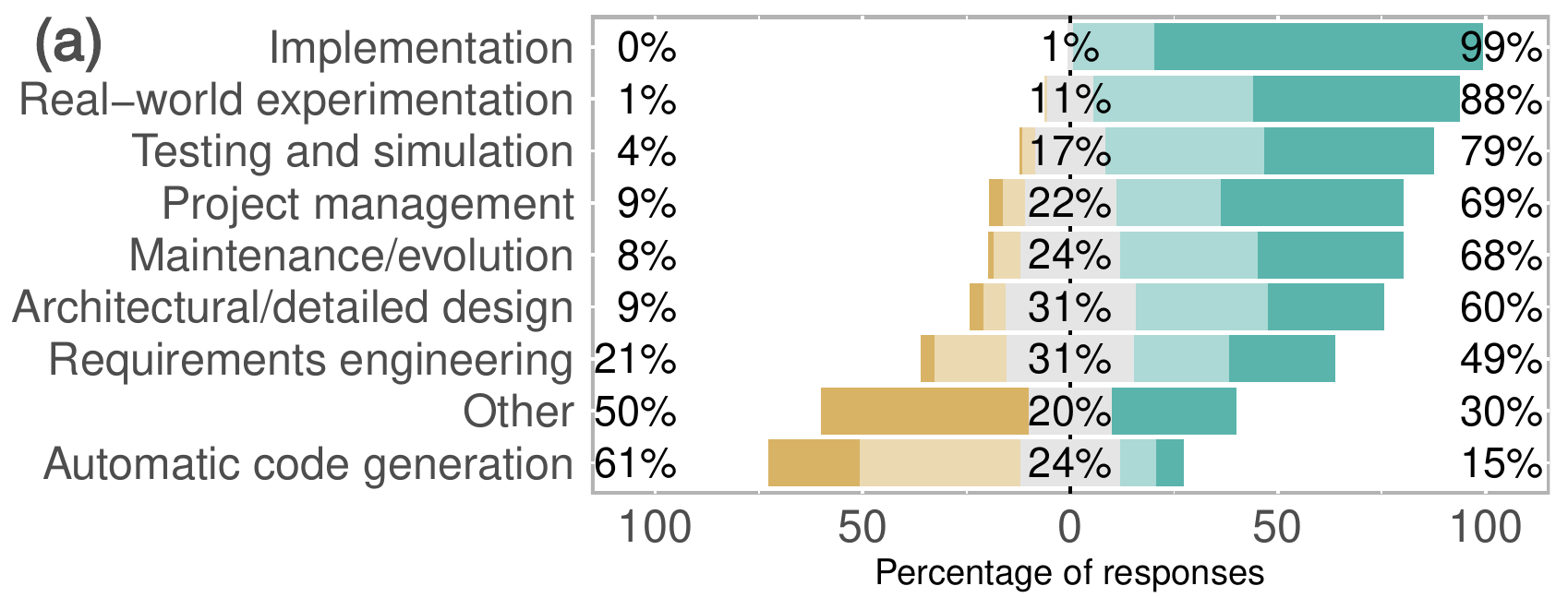}
	\end{minipage}
	\hspace*{\fill}
	\begin{minipage}[t]{0.32\textwidth}
		\includegraphics[
		clip=true, width=\linewidth]{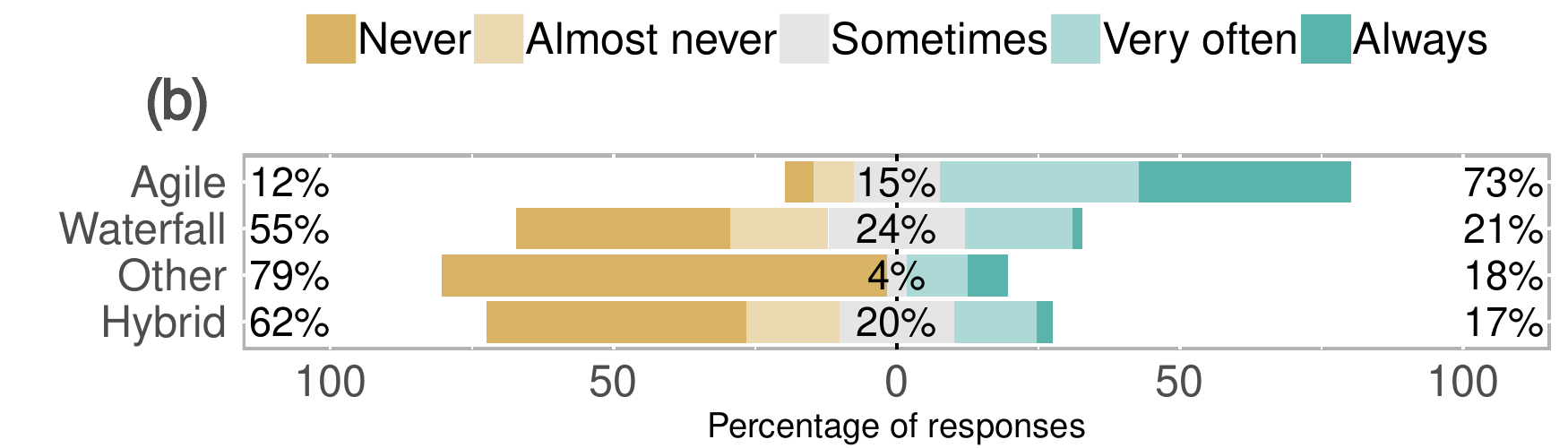}
	\end{minipage}
	\hspace*{\fill}
	\begin{minipage}[t]{0.32\textwidth}
		\includegraphics[clip=true, width=\linewidth]{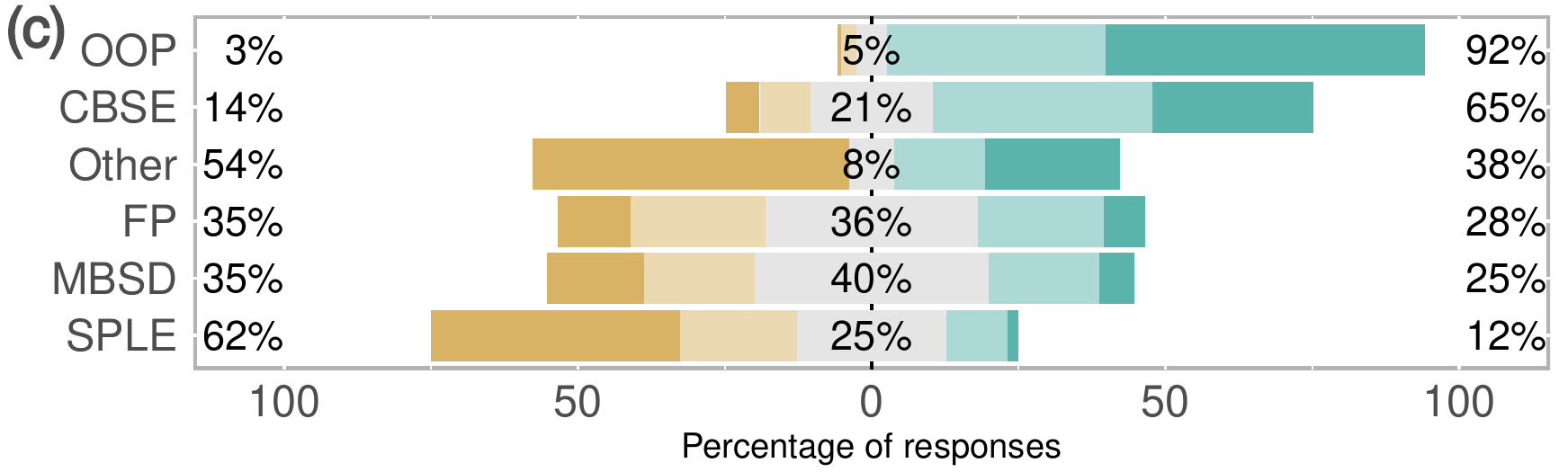}
	\end{minipage}
	\hspace*{\fill}
	
	\vspace{10pt} 
	
	\begin{minipage}[t]{0.32\textwidth}
		\includegraphics[clip=true, width=\linewidth]{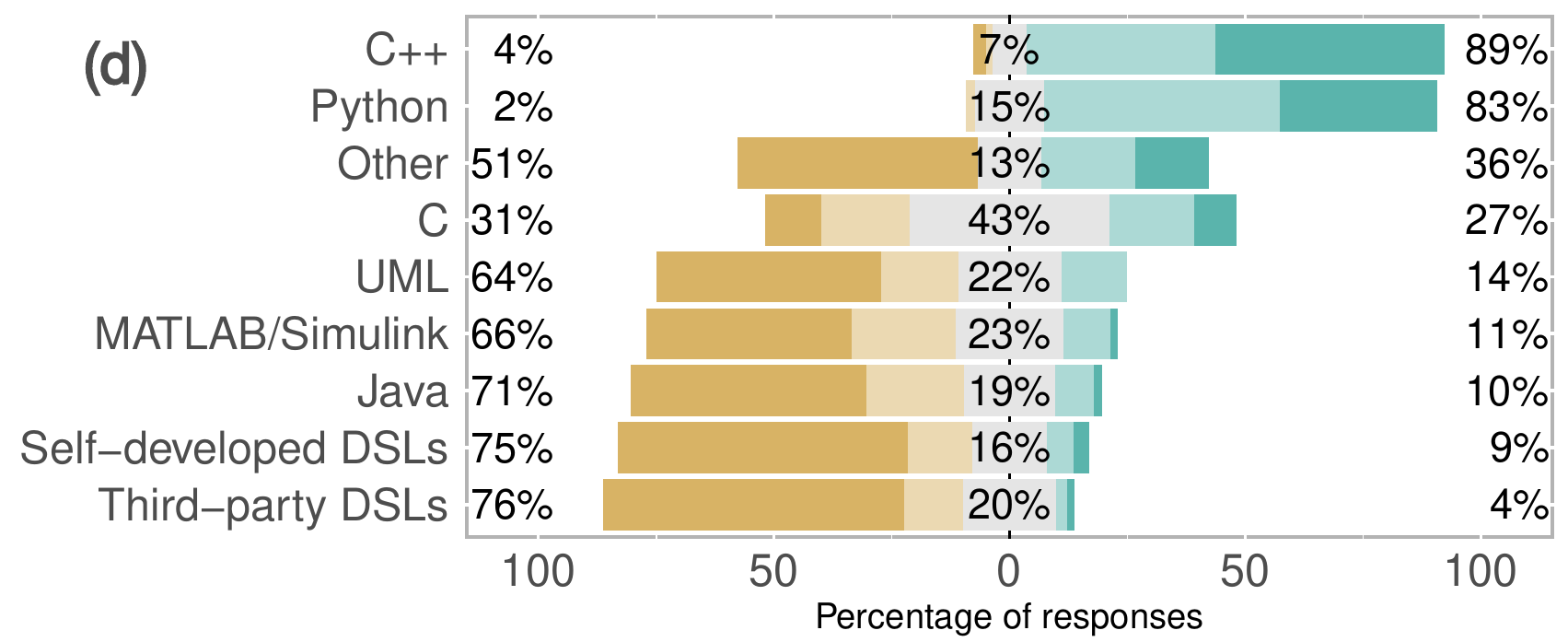}
	\end{minipage}
	\hspace*{\fill}
	\begin{minipage}[t]{0.32\textwidth}
		\includegraphics[clip=true, width=\linewidth]{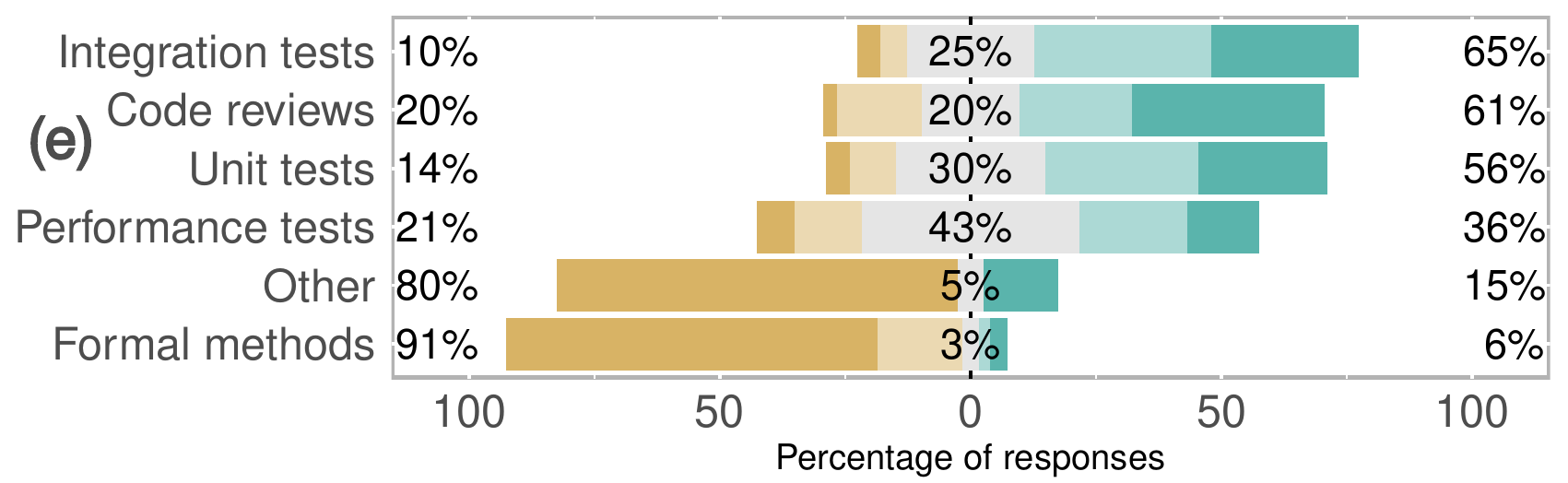}
	\end{minipage}
	\hspace*{\fill}
	\begin{minipage}[t]{0.32\textwidth}
		\includegraphics[clip=true, width=\linewidth]{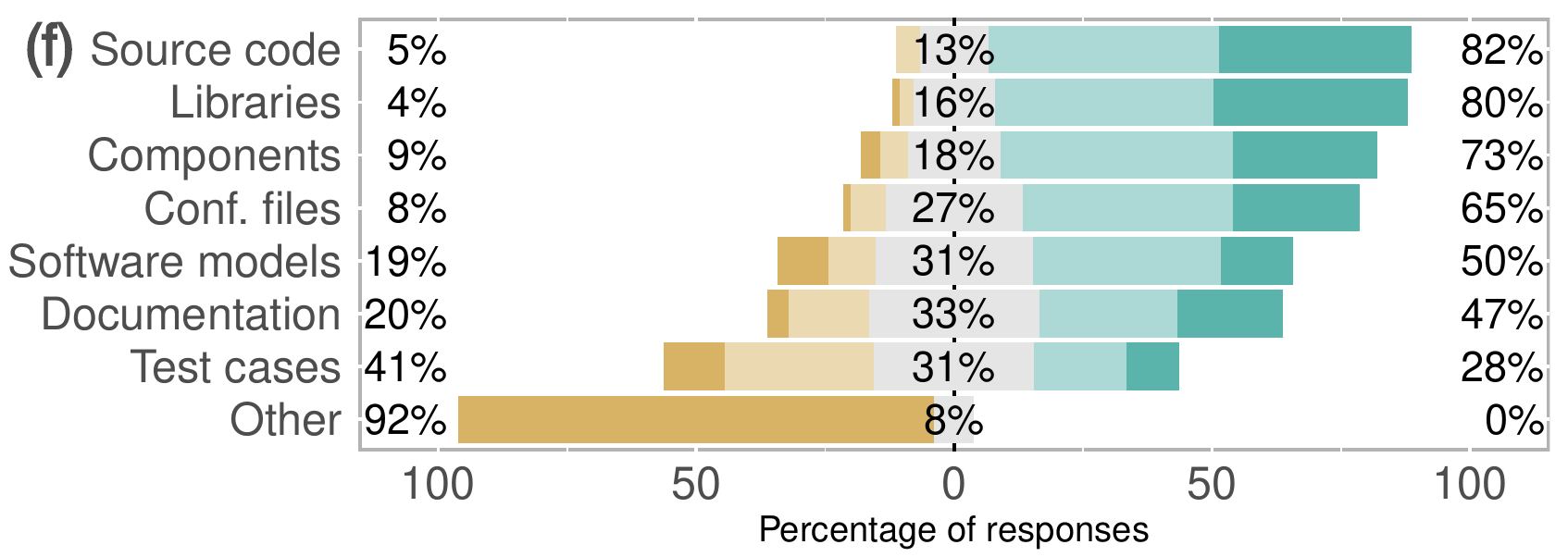}
	\end{minipage}
	\hspace*{\fill}
	\setlength{\abovecaptionskip}{2pt} 
	\caption{Survey Results: Reported Frequencies of (a) Activities; (b) Processes; (c) Paradigms;
		(d) Software Languages; (e) QA Techniques; (f) Reuse of Artifacts.
		The percentages refer to all cases in which the participant specified a frequency. We report on cases with a significant share of ``Don't know'' answers in the text.
	}
	\label{fig:activities} 
	\label{fig:paradigms} 
	\label{fig:processes} 
	\label{fig:testing} 
	\label{fig:languages} 
	\label{fig:artifacts} 
	\vspace{10pt}
\end{figure*}

\section{Practices (RQ1)}
\label{sec:rq1}

RQ1 focuses on practices applied during the software development life cycle of robotic projects. 
It includes typical activities,  development paradigms, development processes, software languages and frameworks, quality assurance (QA), reuse practices, and used tools.


\subsection{Activities} \label{subsec:activities}


\looseness=-1
Figure~\ref{fig:activities} (a) gives an overview of the frequency of different activities according to the survey.
The most common one is implementation, 99\%  of the respondents perform it always or very often.
Real-world experimentation and testing and simulation come in second (88\%) and third place (79\%). 
The least common activities are automatic code generation (15\%) and requirements engineering (RE) with 49\%.

\looseness=-1
Real-world experimentation and testing and simulation are performed equally by industrial and academic practitioners.
Interestingly, respondents perform real-world experimentation more often than simulation (88\% versus 79\%). 
The difference between industrial and academic practitioners for software maintenance and evolution mounts to a rate of 73\% and 56\%, respectively.
The gap may be related to work conventions in academia: academic practitioners do not focus on providing a final product to a customer, but on generating a proof-of-concept design that can be used to validate a scientific contribution.
Only 8\% of the respondents responded ``Don't know'' to the RE question and, for the reminder, the respondents that perform RE are surprisingly low. 
We believe that the respondents had a very specific understanding of RE in the sense of using dedicated RE tools.
Another possible reason is the large number of participants who follow agile processes.  
We analyzed our data and discovered that, from 156 respondents, 70 perform RE, and 48 perform RE \emph{and} agile processes always or very often. 
Therefore, at least a significant portion of respondents who use agile processes indicate RE experience.
Since they all need to somehow elicit requirements, we dug deeper in the interviews and in fact we found that the majority elicit requirements but in rather ad hoc ways. 
\qnorbert{We had the customer requirements and also all the list of features the customer needed, and then we again sat down and decided how to design the software architecture and realize that. But there was no official process that we followed.}
Furthermore, even when they followed an RE process, some stated they did not, probably due to a terminology issue.

\subsection{Development Processes}  \label{subsec:development}


\looseness=-1
We present an overview of the frequency of applied SE processes in Fig.~\ref{fig:processes} (b).
The most common development process applied by the survey respondents is agile (73\% of the respondents apply it always or very often versus 21\% and 17\% for waterfall and hybrid, respectively).
The ``Don't know'' answers' rate is 11\%, 26\%, and 30\%  for agile, waterfall, and hybrid, respectively.
There is a remarkable difference between the industrial and academic practitioners' results: 81\% of the industrial ones adopt agile processes very often or always, while the percentage of academic ones is 60\%.

We observe that agile SE processes are more often followed by industrial practitioners than by their academic counterparts (81\% agains 60\%). 
Three academic practitioners respondents even question how systematic are the processes followed by academia. \qscampusano{Almost always there is no clear SE process}.


\quotebox{\observation{Activities and Development Processes}\label{obs:activities}}{
	Practitioners mainly focus on implementation and real-world experimentation (preferred over simulation) during software development, typically following agile-style processes.
}{0.8cm}
\vspace{-8mm}

\subsection{Paradigms} \label{subsec:paradigms}

\looseness=-1
Figure~\ref{fig:paradigms} (c) gives an overview of the SE paradigms and their use frequency in robotics.
Most survey respondents make use of object-oriented programming (OOP) (92\%) and many follow a CBSE approach (65\%).
Software product line engineering (SPLE) and model-based software development (MBSD) are rarely used (12\% and 25\% use them always or very often, respectively). 
The ``Don't know'' answer's ratio is 32\% and 15\% for SPLE and MBSD, respectively.  


\looseness=-1
The prevalence of OOP and CBSE can be explained with the wide usage of ROS, 
which works with Python and C++ and enforces a component-based approach.
We observe 
that only industrial respondents report frequent use of SPLE, specifically 19\% of the industrial practitioners report using it very often or often, versus 0\% of academic practitioners.
We also conclude that
MBSD is not extensively used in robotics, especially among leading technical role practitioners (only 16\% of them apply it very often or always, in contrast to 30\% and 27\% of industrial programmers and academic practitioners).




\vspace{-10pt}

\subsection{Software Languages and Frameworks}\label{subsec:lang}

Figure~\ref{fig:languages} (d) provides an overview of the use of software languages in robotics.
C++ and Python are used most often; 89\% and 83\%  of our survey respondents use them very often or always, respectively.
Self-developed (9\%) and third-party (4\%) domain-specific languages are the least frequently used among the options. 
The ``Don't know'' answer's rate for C++ and Python is 4\%, while it ranges from 14\% to 23\% for the rest of the options.

Considering frameworks, ROS is undeniably the most used framework for robotics software development nowadays.
In particular, 88.5\% of the survey's respondents and 16 interviewees use it.
ROS2 and OROCOS follow as second and third with 22.4\% and 18.6\%, respectively.
ROS2 is, according to the conducted interviews, not yet implemented in most of the companies, and there is not much initative to change that.
An important reason that drives that decision is stated by \qthilo{Some of our industrial partners are also more conservative. Never change a running system. If you have now ROS1, [...] 
	why change it if it's not clear what exactly  would be the improvement?}



\looseness=-1
For either question, there are no big differences in the results between academic and industrial practitioners, probably due to the widespread use of ROS in both communities.
ROS is mainly written in C++ and Python and provides many libraries that use them.
According to one interviewee,  Python is normally used for high-level components that do not have hard-time constraints or for prototyping, since it is simpler than C++.
\qnavarro{It's really fast and dynamic. [...] For better performance we use C++, and also with C++ we have the possibility to better protect the code than with Python.}
However, a considerable number of the responses (around 20\%) denote a lack of awareness with respect to the other used languages. 


\quotebox{\observation{Paradigms, Languages, and Frameworks}\label{obs:paradigms}}{
	Practitioners usually rely on frameworks to develop software, and ROS is the preferred framework among roboticists.
	Aligned with the popularity of ROS, we find C++ and Python as the most used software languages, and OOP and CBSE as most typical paradigms.
}{0.8cm}
\vspace{-6mm}

\vspace{-10pt}
\subsection{Quality Assurance}\label{subsec:qa}

We investigate quality assurance (QA) techniques and sources of test data for industrial and academic practitioners. 

\parhead{Techniques} Figure~\ref{fig:testing} (e) shows 
that the most common QA technique among respondents is integration testing, 65\% of them apply it very often or always.
Code reviews is the second option (61\%), unit testing the third (56\%), and performance testing is the fourth (36\%).
The least common answer is formal methods (6\%) 
and in fact 43\% answered  ``Don't know''  to this question.

Integration testing may be the most applied QA technique (65\%) due to the popularity of ROS, which enforces a component-based approach for developing software.
Therefore, practitioners need to integrate different components to test a final product. 
Oppositely, industrial practitioners often work in teams.
Surprisingly, performance testing is not performed as often as we expected taking into account the cyber-physical nature of robots and the performance constraints (e.g., regarding memory, power consumption, computation requirements) they entail.

\quotebox{\observation{Quality Assurance Techniques}\label{obs:qa_techniques}}{
	The most widespread QA techniques are code reviews and testing, especially integration testing.
	Surprisingly, despite the role of hardware constraints, performance testing is rarely performed.
}{1.2cm}
\vspace{-5mm}

\looseness=-1
\parhead{Sources of Testing Data}
The survey's results show that the most common source of data for testing is simulation (82.1\%).
The second most used source is runtime monitoring (75\%) and the third manually based on experience (64.7\%).
Testing data is acquired from runtime monitoring (75\%) almost as often as from simulation (82.1\%), which may mean that both industrial and academic practitioners perform almost as much simulation as real-world testing.
This was already identified in Fig.~\ref{fig:activities} (a) and will be further discussed in Sec.~\ref{sec:rq3}.

\quotebox{\observation{Sources of Testing Data}\label{obs:qa_sources}}{
	Practitioners' main source of data is simulation, but the difference with runtime monitoring is surprisingly low.
}{1.6cm}
\vspace{-8mm}

\vspace{-5pt}
\subsection{Reuse}\label{subsec:reuse}

\parhead{Reused Artifacts}
As shown in Fig.~\ref{fig:artifacts} (f), source code and libraries are the most commonly reused artifacts: 82\% and 80\% of the respondents stated that they always or very often reuse them.
Software components and configuration files come in third and fourth with 73\% and 65\%.
The percentages are similar for all the roles concerning these artifacts.
On the other hand, the least reused type of artifact is test cases, which are always or very often reused by 28\% of the respondents.
Among the industrial programmers, 38\% very often or always reuse test cases versus 21\% and 23\% of leading roles and academic practitioners, respectively.

\looseness=-1
Very few respondents reuse documentation (47\% of the respondents reuse it always or very often), in relation to the reuse frequency of source code, libraries, or software components (82\%, 80\%, and 73\%, respectively). 
This finding relates to a challenge we discuss in Sec.~\ref{sec:rq3}: lack of documentation and how it hinders further software reuse (obs.~\ref{obs:standard}).
On the other hand, industrial programmers are the group that most commonly reuses documentation (51\%, 8\% higher than academic practitioners, and 9\% higher than leading roles) and test cases (38\% as opposed to the 28\% of the overall). 
A possible explanation of these findings is that programmers are primarily concerned with understanding, writing, and testing of code.
%

\quotebox{\observation{Reused artifacts}\label{obs:reuse}}{
	Roboticists commonly reuse source code and libraries, but not so often test cases and documentation.
}{2.2cm}
\vspace{-5mm}

%
%

\looseness=-1
\parhead{Roadblocks for reuse}
The main reason (78.2\%) among the survey respondents to not reuse existing software resources relates to technical problems (e.g., component's granularity does not fit, missing functionality, incompatible interfaces).
The second and third most common reasons are lack of documentation (46.8\%) and licensing issues (44.2\%).
Among the respondents, 19.9\% favor self-developed components.
Data regarding technical problems and licensing issues differ between industrial and academic practitioners:
the first group finds more technical problems and licensing issues than the second (83\% versus 69.6\% and 53.2\% versus 28.6\%, respectively).

The second most common cause for not reusing existing software components is lack of documentation (46.8\%), which might be related to the first cause (technical problems with 78.2\%)---i.e., it is hard to assess the specifications of a component without proper documentation.
These findings are in line with literature: According to a recent study~\cite{estefo2019robot},
technical problems are the most common issue (e.g., outdated and buggy resources) for reusing software in ROS---often caused by a lack of maintenance.
The same study points the lack of documentation as a problem for 56\% of respondents.
As expected, licensing issues are more problematic for industrial practitioners than for academic practitioners (53.2\% versus 28.6\%).


\quotebox{\observation{Roadblocks}\label{obs:roadblocks}}{
	The main roadblocks for reusing software artifacts are (mainly) technical problems and lack of documentation.
}{2.2cm}
\vspace{-5mm}

\input{tables/tools.tex}

\parhead{Interoperability}
Remarkably, for interoperability among software components of their robotic systems, practitioners tend to rely on the experience of the team (72.4\% of the responses).
This answer was selected by far more often than those related to a systematic approach to interoperability, such as having a clear definition of the interfaces of each component (42.9\%), following a precise architecture (26.3\%), or following a standardized solution (10.9\%).
The most common standardized solution stated by the survey respondents is ROS itself and its set of standardized messages. 


\quotebox{\observation{Interoperability}\label{obs:interoperability}}{
	To ensure the interoperability among heterogeneous system components, practitioners tend to rely on ad hoc practices.
}{2.2cm}
\vspace{-8mm}



\subsection{Tools}

\looseness=-1
Table~\ref{tab:tools} shows the four most used tools among the respondents for each activity introduced in Fig.~\ref{fig:activities}. 
The use of dedicated tools for project management indicates a high level of commitment to this activity.
Contrary to industrial practitioners, the academic ones favor open-source tools as Redmine (9.9\%) against proprietary ones as Jira (6.9\%).
Respondents (both academic and industrial practitioners) use mostly office suite applications 
for RE.  
This indicates that requirements are usually elicited, specified, and modeled in a simple manner, without using dedicated software.
UML editors (e.g., Papyrus, TikZ-UML, PlantUML), PowerPoint and draw.io are the most common tools for Architectural and Detailed Design. 


The preference of VisualStudio and PyCharm for implementation can be related to the preference of C++ and Python, the two main languages supported by the ROS framework.
Gazebo is the most popular tool for testing and simulation (36\%), which is not surprising considering its strong integration with ROS.
The answers related to the software maintenance and evolution activity are similar to those seen for implementation.


\quotebox{\observation{Tools usage}\label{obs:tools}}{
	Respondents use specialized tools for the most common activities (e.g., Gazebo for simulation) and generic tools for less common ones (e.g., the Office suite for requirements engineering).
}{2.2cm}
\vspace{-5mm}




%% file: tables/tools.tex
\begin{table}[b]
	\vspace{0.2cm}
		{\footnotesize
	\caption{Most used tools for the activities:  Implementation (I), Real-World Experimentation (RWE), Testing/Simulation (TS), Project Management (PM), Software Maintenance/Evolution (SME), Architectural and Detailed Design (ADD), Requirements Engineering (RE), Automatic Code  Generation~(ACG).}
	\vspace{-0.2cm}
	\label{tab:tools}

	\begin{tabular}{ p{0.12cm}  p{1.86cm} p{1.72cm} p{1.43cm} p{1.68cm}  }
		\toprule
		
		\hspace{-0.2cm}\parbox[t]{0.3cm}{\textsf{I}}	
		&
		Visual Studio (22\%)
		&
		QtCreator (10\%)
		&
		Vim (8\%)
		&
		Pycharm (7\%)		
		\\
		
		\hspace{-0.2cm}\parbox[t]{0.3cm}{\textsf{RWE}}	
		&
		Roslaunch (24\%)
		&
		Rosbag (20\%)
		&
		In-house  (11\%)
		&
		ROS (8\%)
		\\
		
		\hspace{-0.2cm}\parbox[t]{0.3cm}{\textsf{TS}}	
		&
		Gazebo (36\%)
		&
		In-house (12\%)
		&
		gtest (8\%)
		&
		Rviz (3\%)
		\\
		
		\hspace{-0.2cm}\parbox[t]{0.3cm}{\textsf{PM}}	
		&
		Jira (19\%)
		&
		Git (18\%)
		&
		Trello	 (14\%)
		&
		Excel (11\%)
		\\

		\hspace{-0.2cm}\parbox[t]{0.3cm}{\textsf{SME}}	
		&
		Visual Studio (20\%)
		&
		Git (18\%)
		&
		Eclipse (7\%)
		&
		Vim (6\%)
		\\
		
		\hspace{-0.2cm}\parbox[t]{0.3cm}{\textsf{ADD}}	
		&
		UML editors (20\%)
		&
		PowerPoint (13\%)
		&
		Draw.io (10\%)
		&
		Confluence (5\%)
		\\
		
		\hspace{-0.2cm}\parbox[t]{0.3cm}{\textsf{RE}}	
		&
		Word (18\%)
		&
		Excel (17\%)
		&
		Jira (14\%)
		&
		Confluence (13\%)
		 \\

		\hspace{-0.2cm}\parbox[t]{0.3cm}{\textsf{ACG}}	
		&  
		In-house (19\%)
		&
		Python (12\%)
		&
		Matlab (8\%)
		&
		None (8\%)
		\\

	\bottomrule

	\end{tabular}
	}
\end{table}

%% file: Sections/rq2.tex
\section{Distinct Characteristics (RQ2)}
\label{sec:rq2}
RQ2 focuses on the distinguishing characteristics of robotics software engineering compared to other domains.
We pay particular attention to the differences to other cyber-physical domains, which are characterized by systems that integrate physical environment and hardware with software---e.g., automotive, avionics. 
We discuss the responses to an open-ended question, in which we asked the survey respondents to name and explain such characteristics.  
We analyzed the respondents' answers to create codes,
group all answers (and relevant parts thereof) using the codes, and counted the matches of answers to codes.
Overall, 156 textual answers were provided and 161 matches counted.
In the following summary, we triangulate the survey results with answers from our interviewees.

\noindent\textbf{Cyber-Physical Systems.}
A total of 64 of our survey respondents perceive robotic systems as cyber-physical systems and 
stress the integration between physical aspects and hardware and software components.
\qscamorga{Any small change in code can be seen easily in the behavior of the robot. On top of that, the work is done really close to the hardware that constitutes the robots.} 
Interactions with the real world lead to a special emphasis on non-functional concerns such as safety and reliability: \qper{I think we have to be very careful when we develop software for systems that are out there in the physical world. If I don't get a particular tweet, that doesn't really matter much. But if my robot goes berserk and starts killing people, that's not good.}

\looseness=-1
\noindent\textbf{Heterogeneous Disciplines.}
Out of the total, 24 survey respondents noted that a robotic system bridges diverse domains such as hardware drivers, firmware, planning, perception, and AI.
\qsbrad{The systems we develop are incredibly complex and involve multiple disciplines and technologies, while doing work in the real world. This means there are dozens of integration points and world interactions that have to be designed, engineered, and tested.}
Our synthesis of the topic is:

\quotebox{\observation{Heterogeneous Disciplines}\label{obs:disciplines}}{
	The creation of a full-working robotic system requires knowledge beyond just software, including hardware, mechanics, and physics.	
}{1.2cm}
\vspace{-5mm}

\noindent On the organizational level, the associated challenges are addressed by engineering teams with diverse experiences and backgrounds.

\begin{figure}[b]
	\vspace{-5mm}
	\begin{center}
		\includegraphics[width=0.8\linewidth]{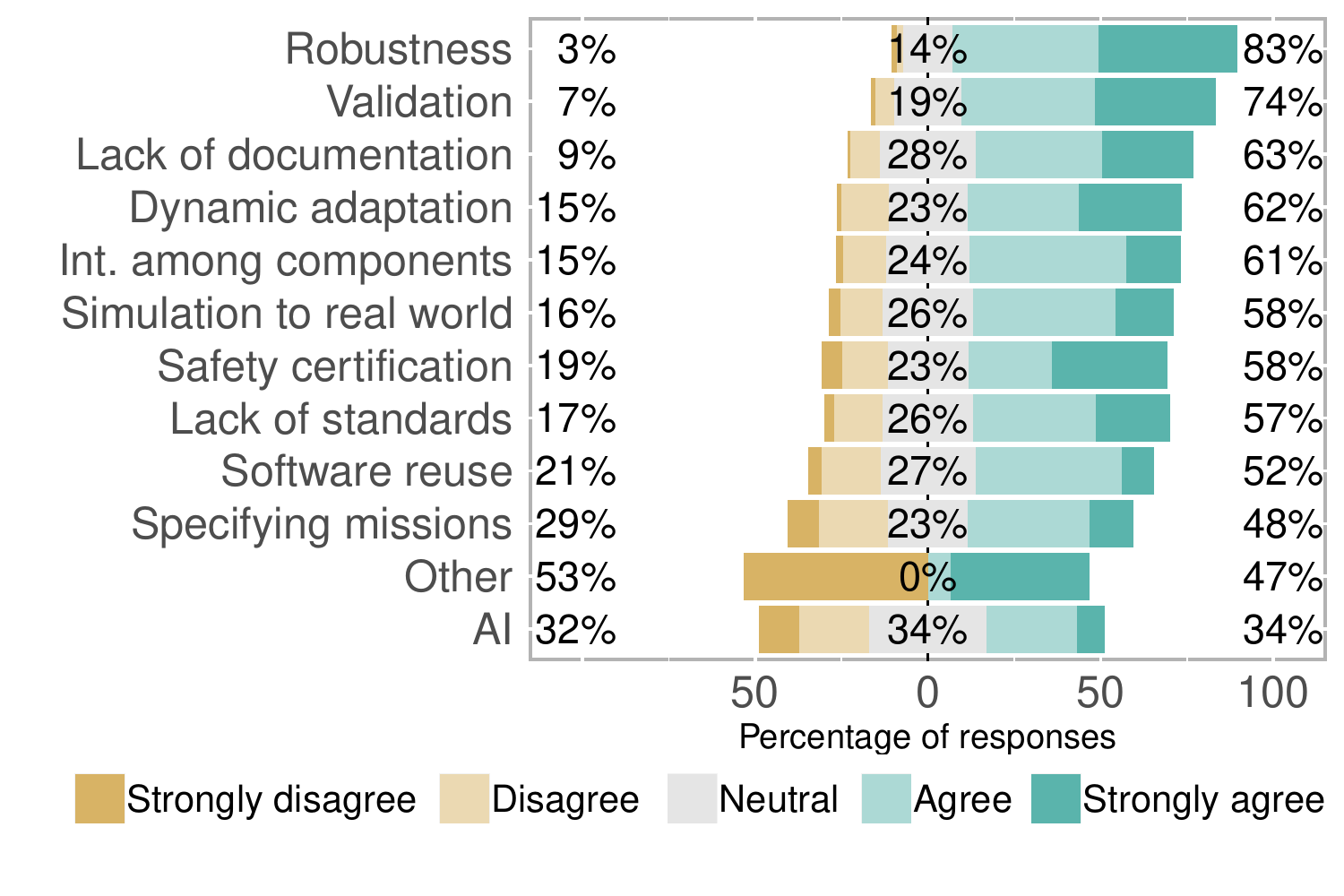}
		\smash{\begin{minipage}{8.5cm}
				\begin{scriptsize}
					\leftline{``Don't know'' (from top to bottom):}
					\leftline{4\%, 7\%, 3\%, 10\%, 8\%, 5\%, 12\%, 4\%, 2\%, 8\%, 90\%, 12\%} 
				\end{scriptsize}
				\vspace{1.85cm}
		\end{minipage}}
		\vspace{-1.1cm}
		\caption{
			Challenges in robotics software engineering (all roles).
		}
		\label{fig:challenges} 
		\vspace{0.15cm}
	\end{center}
\end{figure}

\looseness=-1
\noindent\textbf{Testing.}
As observed by 22 respondents, the interaction with the physical world gives rise to specific requirements on the testing process.
They stress a need for integration testing in the real world. 
\qsdavid{Testing is slower. [It] Requires that you actually have a piece of hardware operating in the real world, where every sensor is lying to you, your environment is actively fighting back, and you have to still maintain sufficient functionality for an outside observer to be happy with your performance. Every assumption must be constantly questioned}.


\looseness=-1
\noindent\textbf{Maturity.}
A total of 20 survey respondents address the maturity of the robotics SE community.
Awareness for SE practices seems to be lower than in other domains.
\qsjulian{Common SE practices are less common or known in robotics and therefore software and software maintainability suffer.}
Our observation is as follows:

\quotebox{\observation{SE practices}\label{obs:maturity}}{
	Awareness of software engineering practices among robotics practitioners is low, which hinders software quality and maintainability.
}{2.2cm}
\vspace{-5mm}

\looseness=-1
\noindent A possible explanation is that software development is often seen as an auxiliary task for developing an actual system.
\qsanonmaturity{Software is not the goal at all. It's only an executable form of robotics theories and algorithms.}
A recurring topic mentioned by three survey respondents is a lack of standardized solutions. 
\qsanonmaturitytwo{In robotics, SE is challenged by the lack of standards and guidelines in the development.}

\noindent\textbf{Context and Adaptation.}
Real-world interactions lead robots to deal with heterogeneous contexts and adaptation, topic that is touched on by 13 survey respondents.
\qsgeorg{Robots act in the real world, which introduces all sorts of unexpected error cases that one wants to catch. Software engineering in other domains often feels <<sandboxy>> in comparison.}
Various responses point out that a robot also becomes an actor that shapes the environment.
\qstahsincankose{They also change the environment itself via actions. Covering all these cases might be intractable in certain scenarios.}

\quotebox{\observation{Adaptable Cyber-Physical Systems}\label{obs:main_characteristics}}{
	Among the most frequently named characteristics of robotics are
	\begin{enumerate*}
		\item their cyber-physical nature,
		\item  the need for testing processes that contemplate the influence from hardware and the physical world, 
		\item and the necessity to react to a large variety of contexts.
	\end{enumerate*} 
}{0.8cm}
\vspace{-5mm}

\noindent\textbf{No difference.}
In contrast to the majority of responses, only eight survey respondents found the differences to other domains negligible. 
One interviewee points out that an emphasis on the differences to other domains can actually lead to a problematic mindset.
\qdavide{It differs much less than people think. [...] Actually, sometimes robotics reinvent the wheel or have a syndrome of <<not invented here>>.} 

\looseness=-1
\noindent\textbf{Influence of academia.}
The influence of academia is commented on by four survey respondents, mostly pointing out a negative influence of academia:
\qsingo{A larger than usual share of code comes from researchers, as opposed to professional software developers. This often causes quality issues.} 


\noindent\textbf{With respect to other CPS.}
What follows are the main differences between robotics and other CPS (e.g., automotive) 
as perceived by four responses:
\begin{enumerate*}
	\item Higher autonomy, more decision making and planning;
	\item  More flexibility for interacting with the world;
	\item  More generic perception capabilities;
	\item More advanced and computationally intensive algorithms (for perception and planning).
\end{enumerate*}

\quotebox{\observation{Distinguishing Characteristics related to other CPS}\label{obs:otherCPS}}{
	The main differences with respect to other CPS are a higher importance of computation requirements and autonomy for decision making, since robots interact with the environment and humans.
}{0cm}
\vspace{-5mm}

%% file: Sections/rq3.tex
\vspace{-0.3cm}
\section{Challenges and Solutions (RQ3)}
\label{sec:rq3}
\looseness=-1
Practitioners in service robotics face various challenges and apply a large variety of solutions to address them. 
In this section, we first present in detail nine of the major challenges that we identified during the first phase of our study (see Fig.~\ref{fig:method}) from the exploratory interviews, our experiences, and literature (specifically, the studies of Brugali~\cite{Brugali2007}, Bures~\cite{Bures2017}, and Garc\'{i}a~\cite{garcia2019variability}).
We quantified the relevance of these challenges in the study’s second phase through the survey.
We report on a rating of the challenges relevance's quantification (see Fig.~\ref{fig:challenges} for an overview), and on their applied solutions as specified in open-ended questions.
Coding 143 responses to the open-ended question yielded 150 solutions; we filtered out 13 answers for being empty or due to not specifying a solution. 





\looseness=-1
\parhead{Robustness}
Most  respondents (83\%) agreed or strongly agreed that the biggest challenge is ensuring enough robustness to robotic systems.
That is, both software and hardware parts that constitute a robot must work continuously without (critical) failures regardless the context---i.e., the mission and the environment.
\qmathias{Programming the [robot's] task is not the most time consuming thing, but programming the failure handling [...] People might need two weeks to set up one application,  two days to set up the task, and eight days to set up all the failure case.} 
We collected 19 solutions for ensuring robotic robustness.
The most frequently named ones are rigorously defined development processes (e.g., V-Model) and thorough testing.

\parhead{Validating Robotic Systems}
It appears to be the second most important challenge, as acknowledged by 74\% of our respondents.
Validation is challenging due to the interaction of robotic systems with the real world, which makes the validation process both expensive and slow (obs.~\ref{obs:main_characteristics}).
We obtained 16 responses with solutions, including 
the employment of an independent team for testing,
defining a standard set of tests the robot or middleware has to be compliant to,
and  formalizing (e.g., modeling) the system as much as possible to automate code generation and validation processes.


\parhead{Lack of Documentation} 
It was considered a major challenge by 63\% of our respondents.
This finding is aligned with our observation from Sec.~\ref{sec:rq1} (obs.~\ref{obs:roadblocks}): 46.8\% of the respondents do not reuse already existing solutions due to lack of documentation.
\qsorin{The biggest challenge for a commercial company is actually to get the right documentation or the right engineers having the right experience and knowledge about those tools.}
In 17 responses with solutions, we find the reviews, use of wikis, guidelines on comments, and automatic generation of documentation as typical solutions.
Five respondents consider this challenge as unsolved, as developers frequently have to understand undocumented source code.



\looseness=-1
\parhead{Dynamic Adaptation} 
It is considered a challenge by 62\% of the respondents.
Dynamic adaptation is becoming an important topic for academic and industrial practitioners, as service robots are expected to adapt to environment conditions~\cite{garcia2019variability}.
This challenge is tightly related with robustness (i.e., autonomy).
From nine  solutions specified by our respondents, 
the most common one (three responses) is to test as many real-world situations as possible, focusing on possible problems and outcomes, which are then modeled and hard-coded  in the system to enable adaptation.
A common modeling paradigm (mentioned by four interviewees) is to define adaptation based on states and pre-defined events, as in state machines or behavior trees.

\quotebox{\observation{Robustness and validation}\label{obs:challengesCPS}}{
	\textit{Achieving robustness}, \textit{validation}, and \textit{dynamic adaptation} are ranked as three of the~top~4 challenges. 
	The main solutions for these challenges are rigorous development and test processes (involving real-world testing) and programming error handling.
}{1cm}
\vspace{-5mm}

\looseness=-1
\parhead{Interoperability} 
This topic is considered a challenge by 61\% of the respondents.
As discussed in Sec.~\ref{subsec:lang} and pointed in obs.~\ref{obs:paradigms}, most robotic projects currently use ROS, which uses a component-based approach.
Therefore, robotic functionalities are often packaged as components that must be integrated into a robotic system.
Integration issues are a consequence of such systems.
From 13 provided responses with solutions, most respondents (seven) suggest to adhere to a clearly defined architecture and to model and document existing components and interfaces.

\looseness=-1
\parhead{Simulation-to-Real-World Transition} 
This issue is considered a major challenge by 58\% of the survey respondents.
Interviewees and respondents claim that current simulation tools are not sufficient to properly validate robotic applications due to issues such as badly-simulated phenomena like physics and sensor feed.
Among 13 collected solutions, five respondents opt for improving the simulation system based on machine learning techniques.
Another five respondents 
suggest transitioning to real-world experimentation as soon as feasible.
\qsandrew{The simulation to real-world gap is real, don't trust any performance measures of anything in simulation, especially if the system is complex. Transition to real world components as soon as it's monetarily feasible.}
Regarding this topic, we found:

\quotebox{\observation{Real World to Simulation Transition}\label{obs:rworld2sim}}{
	Current simulation solutions are not able to emulate real-world phenomena in a sufficiently realistic manner.
	Therefore, practitioners distrust them for validating reliable robotic systems and instead they are relegated to validate basic robotic functionalities.
}{0.8cm}
\vspace{-5mm}




\parhead{Safety Certification} 
More than half of the respondents (58\%)  considered safety certification a major challenge.
As robots interact with the environment, and sometimes with humans, safety becomes a crucial issue.
\qenrique{It's hard to guarantee that a system that relies on learning from a data set, actually generalizes well [...] 
	and learns the right things to be safe, once you deploy it on a completely different environment}.
The 10 solutions obtained for this challenge are rather heterogeneous.
For instance, one respondent employs a specialized team for such issues.
Another respondent partners with a company with expertise in the domain.
A third respondent trains their staff on related standards, such as ISO 13849 and IEC 61508/62061. 


\looseness=-1
\parhead{Lack of Standardized Solutions} 
57\% of the respondents agreed with this topic being a challenge.
According to four of them,  
the lack of standardized solutions in robotics is directly related with its early state of maturity.
\qsivan{The lack of standards, common datasets and problems, and general lack of reuse of software/infrastructure/APIs is making steady progress difficult.}
We collect 14 answers with solutions for this topic.
The one specified the most (named five times) is to develop internal standardized solutions as company guidelines.
Two respondents opt to cherry-pick from solutions from 
other domains (e.g., 
A-SPICE~\cite{AutomotiveSIG2015} from the automotive domain).

\parhead{Software Reuse} 
Among the respondents, 52\%  agreed or strongly agreed with software reuse being a challenge.
\qnorbert{In the future, implementing everything from scratch would get harder and harder. So the challenges that I see are basically how to reuse existing solutions and how to integrate them into your own development.}
We discussed types, artifacts, and roadblocks to reuse in Sec.~\ref{subsec:reuse} (obs.~\ref{obs:reuse},~\ref{obs:roadblocks}).
We obtained 17 responses with solutions to this challenge. 
Six survey respondents encourage modularization and use of libraries and toolboxes.
Other proposed solutions include adherence to standards, use of shared repositories, and model-driven development.

\quotebox{\observation{Lack of Documentation and Established Conventions}\label{obs:standard}}{
	The lack of standard methods, frameworks, libraries, and reference models complicates the development of robotics systems.
	This, together with a large number of existing poorly-documented resources impede interoperability and easy reuse of software, which hinders the development pace of robotics SE.
}{0cm}

\parhead{Specifying Missions}
Almost half of the respondents (48\%) consider mission specification a challenge, being a key issue to make it easy enough for end-users.
Also, mission specification mechanisms should enable users to express ``exceptional" behaviors to specify how robots ought to cope with the {\em variability} of conditions of application scenarios in real environments in which robots are required to operate~\cite{garcia2019variability}.
This is especially problematic in scenarios involving humans (obs 11). 
We collected 10 responses with solutions, divided into two groups: mission specification with and without end-user involvement.
If the end-user is directly involved, methods including DSLs, GUIs, and frameworks are proposed.
If not, developers need to ``hard code'' the mission into the system, typically based on iterative meetings with the end-users.
End-users need to be involved in this iterative mission specification in the case of multi-purpose robots that might be used for a variety of missions and/or scenarios that cannot be or it is impractical to specify upfront.

\parhead{Applying AI Techniques}
This topic was considered a challenge by 34\%  of the respondents.
Out of 18 interviewees, seven apply AI techniques to their projects, five of them to perform obstacle and object detection.
We collected three solutions to this challenge, including applying and extending current research to solve relevant problems, and to provide additional education.

\looseness=-1
\parhead{Others}
Of ``Other'' responses, 90\% answered ``Don't know.'' 
The remaining 10\% provided eight responses, including: ``deterministic execution, timestamping, clock synchronization,'' 
``self-x capabilities,''  and ``lacking extensibility or clear APIs leading to forking.''

%% file: Sections/recommendations.tex
\vspace{-5pt}

\section{Impact and Perspectives}
\label{sec:recommendations}

\looseness=-1
Our collected data and the answers to our three research questions give rise to the following three research themes deserving further attention.
We discuss them along with observations, formulate hypotheses trying to explain them, and propose actions. 

\vspace{-5pt}

\subsection{Robustness and System Validation}


\looseness=-1
\parhead{Observations}
Even though testing practices are among the most widespread quality assurance techniques in robotics (obs.~\ref{obs:qa_techniques}), 
and thorough real-world testing and experimentation processes are mentioned as the most common activities performed by our respondents  (obs.~\ref{obs:activities}),
our results point out that robustness and validation are still perceived as the top challenges of the domain (obs.~\ref{obs:challengesCPS}).
The respondents emphasize the importance of real-world testing especially in the light of the inaccuracy of simulations (obs.~\ref{obs:rworld2sim}, discussed below).
Respondents also highlight the difficulty and the importance of programming robots to handle failures (obs.~\ref{obs:challengesCPS}), i.e., act robustly.


\parhead{Hypotheses}
We hypothesize that the importance of robustness and real-world testing relates to the fact that service robots 
are required to operate without human intervention in a large variety of real-world contexts, which are characterized by an unwieldy diversity of sources of uncertainty.
Systematic testing processes can support the management of possible failures and unexpected environmental events over prolonged timeframes.

\looseness=-1
\parhead{Proposed actions}
As real-world testing is expensive, practitioners would largely benefit from techniques that provide realistic test data.
Most contemporary methods in the testing domain generate test data by relying on code-centric information like coverage criteria and input spaces.
In contrast, there is a potential for novel testing techniques that consider historical real-world data and use them for systematically producing new critical scenarios, which we propose as an action for researchers.
Collections of real-world, industry-based critical scenarios are required to recreate testing cases and therefore improve the robustness of robotic systems.
These critical scenarios should be collected by industrial practitioners.
Moreover, the uncertainty to which robots are exposed asks for novel mechanisms of instilling robustness to robots, e.g. domain-specific languages especially tailored to code environment-based exceptions so to cope with uncertainty.
For the development of such mechanisms, we encourage collaboration among researchers and practitioners, being the former those who take the lead in the idea conception and prototype development and the latter those who apply their knowledge to sharpen the prototype into a refined tool.

\subsection{Simulation}
\parhead{Observations} Since real-world experimentation is time demanding and costly, it is tempting to assume that simulation is the primary means of validating robotic functionalities. 
However, we found that practitioners rely on experimentation as much and sometimes even more than on simulation (obs.~\ref{obs:activities} and ~\ref{obs:qa_sources}).
This is due to the difficulty of modeling real-world phenomena and events typical of environments where robots operate, especially those that involve humans (obs.~\ref{obs:main_characteristics}, \ref{obs:otherCPS}, and \ref{obs:rworld2sim}).
Therefore, improving the capabilities of robotics simulations is a key challenge.


\looseness=-1
\parhead{Hypotheses} The insufficient ability of simulations to validate robotic systems trustworthily can be explained with the hypothesis that existing simulations are decoupled from the real world.
While simulations are typically inspired by real-world phenomena, there is usually little effort to update them with real-world data to make them consistent with real-world observations.
Doing so is a big challenge due to the complexity of the real world (see discussion above).

%

\looseness=-1
\parhead{Proposed actions} Efforts to improve simulation may benefit from
\emph{Multi-Paradigm Modeling }(MPM, \cite{vangheluwe2002introduction}), which suggests the use of appropriate formalisms and abstraction levels to model all relevant aspects
of the system (e.g., the mission to perform, the robot embodiment) and the environment (e.g., human behavior, physics).
The application of appropriate abstraction levels permits to hide the underlying complexity of those aspects, which are modeled with the aim of being usable by domain experts.
Using MPM, those models from different domains are then automatically mapped to a common domain, which can be easily managed by developers~\cite{van2017modelverse}.
Furthermore, MPM has been proposed as an effective manner of tackling the challenges of interdisciplinary domains design as is the example of cyber-physical systems~\cite{van2017modelverse}.
We foresee a huge potential in the application of MPM techniques in robotics to allow domain experts to easily specify their concerns and model specific aspects that can then be integrated into a common domain.
To bridge the gap between the physical and the virtual world in all life cycle phases,
MPM could benefit from the use of \textit{digital twin} technology~\cite{boschert2016digital}, which allows keeping the models synchronized through the development life-cycle based on real-world data.
We recommend close collaborations of researchers and practitioners to develop improved ``live'' MPM frameworks, where the former would act as the lead on the research and development and the latter can help by providing realistic models from their experimentation data.

\subsection{Conventions, Frameworks, and Reference Models}

\looseness=-1
\parhead{Observations}
The development of robotic software mostly consists of integrating reusable components that implement recurrent functionalities. 
While de facto conventions for hardware interfaces in some technical scopes and inter-component communication exist (e.g., ROS messages, see obs.~\ref{obs:paradigms}), 
reusable solutions for recurrent problems in application development are still missing (obs.~\ref{obs:standard}). 
In practice, integration is typically performed in an ad hoc way (obs.~\ref{obs:interoperability}) while being complicated by the wide range of disciplines practitioners working to develop a full-working robotic system have (obs.~\ref{obs:disciplines}).
A related issue is the lack of documentation of implemented solutions, one of the main causes for the lack of reuse of existing resources (obs.~\ref{obs:roadblocks}).
There have been attempts at providing universal solutions for robotics, such as the OMG Robotics Domain Task Force~\cite{omg-robotics}, which aims at fostering the integration of robotic systems.
However, to our knowledge,  the latter has not produced relevant results so far.


\looseness=-1
\parhead{Hypotheses}
We hypothesize that the lack of success of such initiatives comes from the inherent essential complexity of the domain.
Existing conventions on hardware interfaces are not global for the robotics domain, but specific to certain technical scopes, which does not adequately address the continuous supply with new types of hardware. 
Also, the robotics domain as a whole is divided into a variety of sub-domains, including vertical ones (e.g., drivers, planning, navigation) and horizontal ones (e.g., defense, healthcare, logistics).


\looseness=-1
\parhead{Proposed actions}
A common recommendation~\cite{gherardi2014modeling, brugali2012reuse} is to exploit and reuse reference architectures, instead of single components to ease the instantiation, implementation, and reuse of \emph{system abilities}~\cite{mar}.
However, due to the domain's complexity, developing a standardized reference architecture for robotics might be infeasible. 
Instead, we propose to identify a reference architecture for each horizontal domain represented as ``end-user market domains'' in the Multi-Annual Robotics Roadmap~\cite{mar}. 
Yet, these domains are subject to huge variability and it would be required to decompose them into specific, narrow sub-domains that can be used to define reference architectures.
When focusing on a single sub-domain, the reference architecture should be profitably decomposed in various views and viewpoints, as recommended by the ISO/IEC/IEEE 42010 standard~\cite{iso42010}. 
The standard recommends taming the complexity of modern systems through the use of various views, each conforming to a viewpoint, which is identified to frame specific concerns of a precise set of stakeholders.
Then, the various viewpoints are contained into an architecture framework, which defines conventions, principles, and practices for the description of architectures established within a specific domain of application and/or community of stakeholders. 
Two studies of Pelliccione et al. are examples of an architecture framework in the automotive field~\cite{AFVolvo}, and of a specific viewpoint on connected vehicles~\cite{SCP2020}---the latter contains also a functional reference architecture for the specifically connected vehicle viewpoint.

\looseness=-1
Researchers in the software architecture community might bring expertise acquired in other fields to help to establish conventions, principles and practices, and developing architecture frameworks and reference architectures. 
To do so, researchers will need to collaborate with robotics practitioners. 
This may bring an important innovation in the field as it happened in the past with the arrival of middleware frameworks like ROS and OROCOS, which enormously simplified the reuse of software resources by (mainly) establishing communication architectures and interfacing conventions.
This caused a shift in the community mindset, which understood the advantage of those frameworks to solve issues related to system integration---e.g., concurrency, synchronization, real-time communication.
We suggest making similar efforts 
towards the standardization and establishment of conventions, principles, and practices.

%% file: Sections/validity.tex
\vspace{-0.15cm}

\section{Threats to Validity}\label{sec:validity}

We discuss the threats to the validity 
using the standard categorization by Wohlin et al.~\cite{wohlin2012experimentation}.

\looseness=-1
\parhead{External Validity}
To ensure a high degree of external validity, we used a diverse recruitment strategy for our interviews and survey, described in Sec.~\ref{sec:method}.
Our interviewees and respondents work with robotic systems, covering a wide range of domains, contexts, and backgrounds.
The practices, characteristics, and challenges we describe are applicable to service robots in similar domains. 

\looseness=-1
\parhead{Construct Validity}
The main threat are terminology issues:
while we benefit from our own experience in the domain, a lack of a shared understanding with our respondents may threaten our conclusions.
As one example, we point out the case of requirements engineering (RE): only 49\% of all survey respondents specify that RE is performed very often or always in their organizations. We noticed that several interviewees had a somewhat narrow understanding of RE, focusing on specialized notation and tools, and that a RE process in a wider sense was universally applied.
Apart from this case, by collecting and triangulating qualitative and quantitative data, we were able to observe that the general trends were in line in both data sources.

\parhead{Internal Validity}
We relied on voluntary participants from industry, most of whom we did not know personally.
Therefore, our sample may be affected by selection bias:
it may be skewed towards participants who are willing to participate in volunteering activities, which may be correlated with some relevant phenomena such as the willingness to contribute to open-source projects.

\parhead{Conclusion Validity}
Coding as we performed it for our qualitative analysis is subject to interpretation. 
We mitigated bias by cross-checking the obtained codes and refining the coding in a collaborative manner.
We triangulated the coding with our qualitative and quantitative data to enhance our study's validity.

%% file: Sections/conclusion-rec.tex
\vspace{-0.1cm}

\section{Conclusion}
\label{sec:conclusion}
\looseness=-1
We presented an empirical assessment of the current state of robotics SE, comprising 18 interviews with industrial practitioners from 15 different companies and a survey with 156 respondents from 26 countries. 
We discussed
\begin{enumerate*}
	\item common practices applied by practitioners for developing robotic applications,
	\item the main characteristics of robotics SE, and
	\item challenges commonly faced by practitioners and applied solutions.
\end{enumerate*}
Robotic systems are required to provide higher levels of autonomy in terms of decision making and planning because they must interact with (and react to) the environment and humans.
Software development for such systems renders important challenges and calls for systematic SE approaches 
to build them.
During our study, we found an overall lack of SE best practices   
in robotics development 
(obs.~\ref{obs:maturity}) and a substantial awareness of this issue. 
In our understanding, robotics would enormously benefit from the application of such practices to solve some of the challenges we elicited---e.g., software architectures to cope with software and hardware interoperability, modeling to promote modularity and reusability.
We hope our work contributes to raising awareness of the importance of further investigation of this topic.


\looseness=-1
Based on our discussion, possible directions for future work are:
\begin{enumerate*}
	\item Investigating novel testing techniques that consider real-world data;
	\item the development of ``live'' multi-paradigm modeling frameworks that use the digital twin technology; and
	\item the conception of reference architectures for specific end-user market domains.
\end{enumerate*}


